\documentclass[aps,graphicx,12pt,showkeys,showpacs,onecolumn]{revtex4}
\usepackage{amsmath}
\usepackage{amscd}
\usepackage{graphicx}
\usepackage{subfigure}
\usepackage{appendix}

\begin{document}

\title[Short Title]{Shortcuts to adiabatic passage for multiparticle in distant cavities:
                    Applications to fast and noise-resistant quantum population transer, entangled states' preparation and transition}

\author{Ye-Hong Chen$^{1}$}
\author{Yan Xia$^{1,}$\footnote{E-mail: xia-208@163.com}}
\author{Qing-Qin Chen$^{2}$}
\author{Jie Song$^{3}$}

\affiliation{$^{1}$Department of Physics, Fuzhou University, Fuzhou
350002, China\\$^{2}$Zhicheng College, Fuzhou University, Fuzhou
350002, China\\$^{3}$Department of Physics, Harbin Institute of
Technology, Harbin 150001, China}


\begin{abstract}

  In this paper, we study the fast and noise-resistant population
  transfer, quantum entangled states preparation, and quantum entangled
  states' transition by constructing the shortcuts to adiabatic
  passage (STAP) for multiparticle based on the approach of ``Lewis-Riesenfeld
  invariants'' in distant cavity quantum electronic
  dynamics (QED) system. Numerical simulation demonstrates that all
  of the schemes are fast and robust against the
  decoherence caused by atomic spontaneous emission and photon leakage.
  Moreover, not only the total operation time but also the
  robustness in each scheme against decoherence is irrelevant to the
  number of qubits. This might lead to a useful step toward realizing the fast and noise-resistant quantum information
  processing in current technology.

\end{abstract}


\maketitle
\section{Introduction}
Quantum information processing (QIP) has demonstrated an important
development in recent years. A crucial prerequisite for QIP is the
ability to generate and manipulate various highly nonclassical and
entangled states which are fundamental for demonstrating quantum
nonlocality \cite{JSBPhys65,DMGMAHASAZAjp90}. There are two widely
used methods for generating and manipulating the entangled states
with external interacting fields. One is fixed-area resonant pulses
route \cite{LAJHENy87,ECGIAAPJDVDEPrl04,BTTSGNVVPrl11}, and the
other one is adiabatic method
\cite{KBHTBWSRmp98,NVVTHBWSKBArpc01,PKITMSRmp07} in which total
Hamiltonian $H (t)$ depends explicitly on time. Generally speaking,
to drive the system evolution with the simple time-independent fixed
area resonant pulses may be fast, but the schemes are difficult to
implement because all parameters need to be informed and the
interaction time is required to be controlled exactly. Moreover, the
fidelities are highly sensitive to the fluctuations of parameters.
The adiabatic passage is robust to the fluctuations of parameters,
whereas, a long evolution time is demanded. In other words, a slow
change of $H (t)$ with time
guarantees each instantaneous eigenstates of $H (t)$ evolves along
itself all the time without converting to other states and ensures
that the adiabatic passage technique is realized. Otherwise, the
evolution of the system would be unpredictable and the fidelities of
the target states could be degraded. However, in a real situation,
when the required evolution time is too long, the method will be
useless because the dissipation caused by decoherence, noise, and
losses on the target state increases with the increasing of the
interaction time.
There are many instances that we would like or need to quicken the
operations in experiment. An idea method to generate and manipulate
various entangled states should be fast, robust and easy realized
with the current technology. Combining the best of the resonant
pulses route and adiabatic method, accelerating the dynamics of
adiabatic passage towards the final outcomes is a perfect way to
realize fast and noise-resistant population transfer, quantum
entangled states preparation, and quantum entangled states
transition under the current technology.

``Shortcuts to adiabatic passage'' (STAP) \cite{XCILARDGJGMPrl10,ETSLSMGMMADCDGOARXCJGMAmop13,ARXCDAJGMNjp12} which is recently and timely introduced to describe
schemes that speed up a quantum adiabatic process usually, although
not necessarily, through a non-adiabatic route, has attracted a
great deal of attention and promises to overcome the harmful effect
caused by decoherence, noise or losses during a long operation time.
In recent years, various reliable, fast, and robust schemes have
been proposed in finding shortcuts to an slow adiabatic passage in
theory
\cite{XCILARDGJGMPrl10,ARXCDAJGMNjp12,AdCPra11,XCETJGMPra11,XCJGMPra12,MDSARJpca03Jpcb05Jcp08,MVBJpamt09} 
 and 
 in experiment
 \cite
{PRA-82-033430-2010,EPL-93-23001-2011,PRL-109-080501-2012,
PRL-109-080502-2012,OL-37-5118-2012,Nature-8-147-2012}. For example,
Chen \emph{et al.} \cite{XCILARDGJGMPrl10} have put forward a scheme
to speed up the adiabatic passage techniques in two- and three-level
atoms. Later, they have also proposed a scheme \cite{XCJGMPra12} to
perform fast population transfer (FPT) in three-level systems with
the help of the invariant-based inverse engineering and the resonant
laser pulses. However, it worth noticing that although many methods
for STAP have been realized in the internal states of a single atom
in different systems, it is very hard to directly generalize these
ideas to two- and multi-particle cases. In view of that we are led
to ask if it is possible to construct STAP for multiparticle
systems. In this scenario, Lu \emph{et al.} \cite{MLYXLTSJSNBAPra14}
have proposed a scheme to implement the FPT and fast maximum
entanglement preparation between two atoms in a cavity QED based on
the transitionless quantum driving proposed by Berry
\cite{MVBJpamt09}. Then, Lu \emph{et al.} \cite{MLLTSYXJS13} have
also put forward another scheme to realize the fast quantum sate
transfer between two three-level atoms using the invariant-based
inverse engineering in a cavity QED.  In 2014, motivated by the
quantum Zeno dynamics, Chen \emph{et al.} \cite{YHCYXQQCJSPRA14}
have constructed shortcuts for performing the FPTs of ground states
in multiparticle systems with the invariant-based inverse
engineering \cite{XCJGMPra12}. This scheme is not only implemented
without requiring extra complex conditions, but also insensitive to
variations of the parameters.

We note that refs.
\cite{MLYXLTSJSNBAPra14,MLLTSYXJS13,YHCYXQQCJSPRA14,XSLFWarxiv13}
have successfully introduced STAP into cavity QED systems which concern
the interaction of atoms and photons within a cavity and are very
promising and highly inventive for QIP
\cite{JMRMBSHRmp01,MASetalNat07,JMetalNat07,HJKNat08}. However, they
\cite{MLYXLTSJSNBAPra14,MLLTSYXJS13,YHCYXQQCJSPRA14,XSLFWarxiv13}
all assume the case that all the operations are implemented in only
one place, for example, a cavity QED. In view of the requirements
for long-distant quantum computation and quantum information
processing, it is desirable to extend the approach to distant
cavities system. But, when it comes to more complex systems, for
example, multi-cavity-fiber-atom combined system, the above schemes
\cite{MLYXLTSJSNBAPra14,MLLTSYXJS13,YHCYXQQCJSPRA14,XSLFWarxiv13}
are useless. That is, new designs are required in a complex
situation. In fact, constructing STAP for multiparticle in
cavity-fiber-atom combined system is very complicated since it is
difficult to look for a Hamiltonian operator $I(t)$
($i\partial_{t}I(t)=[H(t),I(t)]$) that is related to the original
Hamiltonian $H(t)$ but drives the eigenstates
$\{|\Psi_n(t)\rangle\}$ exactly. Therefore, until now, this problem
has not been addressed.

In this paper, we use the approach of ``Lewis-Riesenfeld (LR)
invariants'' to construct shortcuts to 
speed up the rate of the quantum population transfer, the quantum
entangled states generation, and the quantum entangled states
transition for multiparticle in multi-cavity-fiber-atom combined
system. That is, we first study how to construct STAP by
inverse engineering for $N$ ($N \in \{1,2,3,\cdots ,+\infty\}$)
atoms which are trapped in $N$ distant optical
cavities, respectively. 
Then we use STAP to realize the fast and noise-resistant
quantum population transfer, Bell states \cite{bell},
Greenberger-Horne-Zeilinger (GHZ) states \cite{DMGMHASAZAjp90} and
$W$ states \cite{WDGVJICPra00} preparation, and entangled states
transition
\cite{CHBGBCCPrl93,BSSYKJGCGPla00,JICPZHJKHMPrl97,ZQYFLLPRA07,SBZGCGPrl00}
in spatially separated cavities which are connected by fibers.
Compared with previous works, the present schemes have the following
advantages. First, the FPT, the fast quantum entangled states
generation, and the fast quantum entangled states transition for
multiparticle in spatially separated atoms can be achieved in one
step. Secondly, our schemes are not only fast, but also robust
versus variations in the experimental parameters and decoherence
caused by atomic spontaneous emission and photon leakage. In fact,
further research shows that, both the total operation time and the
robustness against decoherence in each of the schemes is irrelevant
to the number of qubits.

The paper is structured as follows. In Sec. \textrm{II}, we briefly
describe the LR phases. In Sec. \textrm{III}, we construct STAP
for FPT in a system with two spatially separated atoms trapped in
different cavities which are connected by a fiber through designing
resonant time-dependent laser pulses by invariant-based inverse
engineering. In Sec. \textrm{IV}, we use STAP to fast
prepare Bell state, GHZ state, and $W$ state in two distant cavities
which are connected by a fiber. In Sec. \textrm{V} we generalize the
schemes in Sec. \textrm{IV} to implement the fast entangled state
transition. In Sec. \textrm{VI}, we give the numerical simulation
and discussion for our schemes. The conclusion appears in Sec.
\textrm{VII}.

\section{Lewis-Riesenfeld phases}
We would like to give a brief description about the LR theory
\cite{HRLWBRJmp69,MALJpa09}. We consider a time-dependent quantum
system whose Hamiltonian is $H(t)$. Associated with the Hamiltonian
there are time-dependent Hermitian invariants of motion $I(t)$ that
satisfy
\begin{eqnarray}\label{eq2-1}
  i\hbar\frac{\partial I(t)}{\partial t}-[H(t),I(t)]=0.
\end{eqnarray}
For any solution $|\Psi(t)\rangle$ of the time-dependent
Schr\"{o}dinger equation $i\hbar
\partial_{t}|\Psi(t)\rangle=H(t)|\Psi(t)\rangle$ ($\partial_{t}=\frac{\partial}{\partial t}$), $I|\Psi(t)\rangle$
is also a solution, and $|\Psi(t)\rangle$ can be expressed as a
linear combination of invariant modes
\begin{eqnarray}\label{eq2-2}
  |\Psi(t)\rangle=\sum_{n}C_{n}e^{i\alpha_{n}}|\phi_{n}(t)\rangle,
\end{eqnarray}
where $C_{n}$ is the $n$th constant, $|\phi_{n}(t)\rangle$ is the
$n$th eigenvector of $I(t)$ and the corresponding real eigenvalue is
$\lambda_{n}$. The LR phases $\alpha_{n}$ fulfill
\begin{eqnarray}\label{eq2-3}
  \hbar\frac{d\alpha_{n}}{dt}=\langle\phi_{n}(t)|i\hbar\frac{\partial}{\partial t}-H(t)|\phi_{n}(t)\rangle.
\end{eqnarray}

\section{Fast population transfer in two spatially separated atoms}

\begin{figure}
 \scalebox{0.25}{\includegraphics {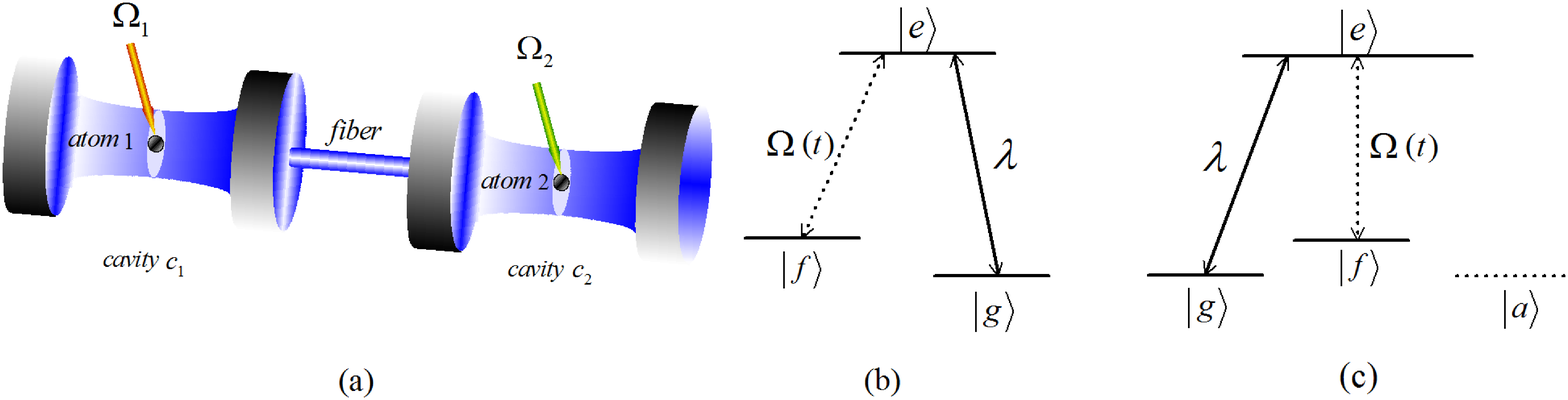}}
 \caption{ (a) The set-up diagram. (b) The atomic level configuration for the $\Lambda$-type atom.
           (c) The atomic level configuration for the atoms with the auxiliary ground state $|a\rangle$ in the schemes of Bell-state and GHZ-state generations.}
 \label{model}
\end{figure}

For the sake of the clearness, as shown in Figs. \ref{model} (a)
and (b), we first assume that two ($N=2$) $\Lambda$-type atoms are
trapped in two distant optical cavities $c_{1}$ and $c_{2}$ which
are connected by a fiber, respectively. Each atom has an excited
state $|e\rangle$ and two ground states $|f\rangle$ and $|g\rangle$.
The atomic transition $|f\rangle_{k}\leftrightarrow|e\rangle_{k}$
($k=1, 2$) is resonantly driven by the $k$th laser pulse with the
time-dependent Rabi frequency $\Omega_{k}(t)$ and the transition
$|g\rangle_{k}\leftrightarrow|e\rangle_{k}$ resonantly couples to
the $k$th cavity mode with the ordinary coupling constant
$\lambda_{k}$.
$\Omega_{k}(t)$ and $\lambda_{k}$ are assumed to be
real in the following for simplicity. In the short-fiber limit,
$L\tau/(2\pi c)\ll1$ \cite{TPPrl97,ASSMSBPrl06}, where $L$ denotes
the fiber length, $c$ denotes the speed of light, and $\tau$ denotes
the decay of the cavity field into a continuum of fiber mode, only
one resonant fiber mode interacts with the cavity mode. The
Hamiltonian of the whole system in the interaction picture can be
written as ($\hbar=1$)
\begin{eqnarray}\label{eq3-1}
  H_{I} &=&H_{al}+H_{ac}+H_{cf},                                          \cr\cr
  H_{al}&=&\sum_{k=1,2}{\Omega_{k}(t)|e\rangle_{k}\langle f|}+H.c.,       \cr\cr
  H_{ac}&=&\sum_{k=1,2}{\lambda_{k}|e\rangle_{k}\langle g|a_{k}}+H.c.,    \cr\cr
  H_{cf}&=&vb_{f}^{\dag}(a_{1}+a_{2})+H.c.,
\end{eqnarray}
where $a_{k}$ is the annihilation operator for the $k$th cavity
mode, $b_{f}^{\dag}$ is the creation operator for the fiber, and $v$
is the coupling constant between the fiber and the cavites. We
assume the initial state is
$|\psi_{0}\rangle=|f\rangle_{1}|g\rangle_{2}|0\rangle_{c_{1}}|0\rangle_{c_{2}}|0\rangle_{f}$,
the whole system evolves in the subspace spanned by
\begin{eqnarray}\label{eq3-2}
  |\psi_{1}\rangle&=&|f\rangle_{1}|g\rangle_{2}|0\rangle_{c_{1}}|0\rangle_{c_{2}}|0\rangle_{f}, \cr
  |\psi_{2}\rangle&=&|e\rangle_{1}|g\rangle_{2}|0\rangle_{c_{1}}|0\rangle_{c_{2}}|0\rangle_{f}, \cr
  |\psi_{3}\rangle&=&|g\rangle_{1}|g\rangle_{2}|1\rangle_{c_{1}}|0\rangle_{c_{2}}|0\rangle_{f}, \cr
  |\psi_{4}\rangle&=&|g\rangle_{1}|g\rangle_{2}|0\rangle_{c_{1}}|0\rangle_{c_{2}}|1\rangle_{f}, \cr
  |\psi_{5}\rangle&=&|g\rangle_{1}|g\rangle_{2}|0\rangle_{c_{1}}|1\rangle_{c_{2}}|0\rangle_{f}, \cr
  |\psi_{6}\rangle&=&|g\rangle_{1}|e\rangle_{2}|0\rangle_{c_{1}}|0\rangle_{c_{2}}|0\rangle_{f}, \cr
  |\psi_{7}\rangle&=&|g\rangle_{1}|f\rangle_{2}|0\rangle_{c_{1}}|0\rangle_{c_{2}}|0\rangle_{f}.
\end{eqnarray}
The states $|\psi_{2}\rangle-|\psi_{6}\rangle$ can be regarded as
``intermediate'' states when we transfer the population from the
state $|\psi_{1}\rangle$ to $|\psi_{7}\rangle$. Hence, we regard the
Hamiltonian $H_{im}=H_{ac}+H_{cf}$ as an ``intermediate''
Hamiltonian. Then we rewrite the above Hamiltonian in Eq.
(\ref{eq2-1}) with a set of vectors $\{|\psi_{1}\rangle,\
|\psi_{7}\rangle,\ |\phi_{0}\rangle,\
|\mu_{1}\rangle=(|\phi_{1}\rangle+|\phi_{2}\rangle)/\sqrt{2},\
|\mu_{2}\rangle=(|\phi_{1}\rangle-|\phi_{2}\rangle)/\sqrt{2},\
|\mu_{3}\rangle=(|\phi_{3}\rangle+|\phi_{4}\rangle)/\sqrt{2},\
|\mu_{4}\rangle=(|\phi_{3}\rangle-|\phi_{4}\rangle)/\sqrt{2}\}$,
where
\begin{eqnarray}\label{eq3-3}
  |\phi_{0}\rangle&=&\frac{v}{\sqrt{2v^{2}+\lambda^{2}}}(|\psi_{2}\rangle-\frac{\lambda}{v}|\psi_{4}\rangle+|\psi_{6}\rangle),    \cr
  |\phi_{1}\rangle&=&\frac{1}{2}(-|\psi_{2}\rangle-|\psi_{3}\rangle+|\psi_{5}\rangle+|\psi_{6}\rangle),                           \cr
  |\phi_{2}\rangle&=&\frac{1}{2}(-|\psi_{2}\rangle+|\psi_{3}\rangle-|\psi_{5}\rangle+|\psi_{6}\rangle),                           \cr
  |\phi_{3}\rangle&=&\frac{\lambda}{2\sqrt{2v^{2}+\lambda^{2}}}
                     (|\psi_{2}\rangle
                     +\frac{\sqrt{2v^{2}+\lambda^{2}}}{\lambda}|\psi_{3}\rangle
                     +\frac{2v}{\lambda}|\psi_{4}\rangle
                     +\frac{\sqrt{2v^{2}+\lambda^{2}}}{\lambda}|\psi_{5}\rangle
                     +|\psi_{6}\rangle),                                                                                          \cr
  |\phi_{4}\rangle&=&\frac{\lambda}{2\sqrt{2v^{2}+\lambda^{2}}}
                     (|\psi_{2}\rangle
                     -\frac{\sqrt{2v^{2}+\lambda^{2}}}{\lambda}|\psi_{3}\rangle
                     +\frac{2v}{\lambda}|\psi_{4}\rangle
                     -\frac{\sqrt{2v^{2}+\lambda^{2}}}{\lambda}|\psi_{5}\rangle
                     +|\psi_{6}\rangle),
\end{eqnarray}
are the eigenvectors of $H_{im}$ corresponding eigenvalues
$E_{0}=0$, $E_{1}=\lambda$, $E_{2}=-\lambda$,
$E_{3}=\chi=\sqrt{2v^{2}+\lambda^{2}}$ and
$E_{4}=-\chi=-\sqrt{2v^{2}+\lambda^{2}}$. And we obtain
\begin{eqnarray}\label{eq3-4}
  H_{re}&=&\frac{v}{\chi}|\phi_{0}\rangle(\Omega_{1}\langle\psi_{1}|+\Omega_{2}\langle\psi_{7}|)
           +\frac{1}{\sqrt{2}}|\mu_{1}\rangle(-\Omega_{1}\langle\psi_{1}|+\Omega_{2}\langle\psi_{7}|)         \cr\cr
           &&+\frac{\lambda}{\sqrt{2}\chi}|\mu_{3}\rangle(\Omega_{1}\langle\psi_{1}|+\Omega_{2}\langle\psi_{7}|)
           +\lambda|\mu_{1}\rangle\langle\mu_{2}|+\chi|\mu_{3}\rangle\langle\mu_{4}|+H.c..
\end{eqnarray}
Through solving the eigenvalue equation of $H_{re}$, the
instantaneous dark state is given by:
\begin{eqnarray}\label{eq3-5}
  |D\rangle=\frac{1}{N_{D}}(\Omega_{2}|\psi_{1}\rangle-\Omega_{1}|\psi_{7}\rangle+\frac{\sqrt{2}\Omega_{1}\Omega_{2}}{\lambda}|\mu_{2}\rangle).
\end{eqnarray}
In light of adiabatic process, we know when the adiabatic
condition $\langle \theta_{0}|\partial_{t}\theta_{m\neq0}\rangle\ll
\xi_{m}$ is fulfilled, where $|\theta_{l}\rangle$ and $\xi_{l}$
($l=0,1,\cdots,6$) denote the instantaneous eigenstates and
eigenvalues of $H_{re}$, respectively, the initial state undergoes
the evolution decided by Eq. (\ref{eq3-5}). However, the simplest
way of speeding up the evolution is to construct nonadiabatic
processes for the system. The states $|\theta_{m\neq0}\rangle$ are
no longer completely negligible but required for the STAP process.
Actually, most of the eigenstates are still completely negligible
($\langle \theta_{0}|\partial_{t}\theta_{m\neq0}\rangle\ll \xi_{m}$
is fulfilled), only $|\theta_{1}\rangle$ and $|\theta_{2}\rangle$
whose eigenvalues $\xi_{1}$ and $\xi_{2}$ are closest to zero have
the chance to participate in the evolution, otherwise, the dynamic
is far different from adiabatic passage that we can no longer name
it ``shortcuts to adiabatic passage''. To speed up the transfer by
using the dynamics of invariant based inverse engineering, we need
to introduce an invariant Hermitian operator $I(t)$ which satisfies
$i\partial_{t}I(t)=[H_{re},I]$
\cite{XCILARDGJGMPrl10,XCETJGMPra11,YHCYXQQCJSPRA14,HRLWBRJmp69},
whereas, it is obvious that directly designing such an operator for
$H_{re}$ is full of challenges. Therefore, some simplifications
should be taken first. With the help of Zeno space division
\cite{PFSPPrl02}, we partition the ``intermediate'' states (the
terms governed by $H_{im}$) into three parts which are independent
of each other:
\begin{eqnarray}\label{eq3-6}
  S_{0}=\{|\phi_{0}\rangle\}, S_{1}=\{|\mu_{1}\rangle,|\mu_{2}\rangle\}, S_{2}=\{|\mu_{3}\rangle,|\mu_{4}\rangle\}.
\end{eqnarray}
One can find from Eq. (\ref{eq3-4}) that the states
$|\mu_{2}\rangle$ and $|\mu_{4}\rangle$ only can be transformed from
the states $|\mu_{1}\rangle$ and $|\mu_{3}\rangle$, respectively. If
we set limiting conditions to make the states $|\mu_{1}\rangle$ and
$|\mu_{3}\rangle$ negligible during the evolution, the states
$|\mu_{2}\rangle$ and $|\mu_{4}\rangle$ become independent from the
system. To neglect the state $|\mu_{4}\rangle$, we regard the
Hamiltonian $H_{re}$ as $H_{re}=H_{1}+H_{0}$, where
$H_{0}=\chi|\phi_{3}\rangle\langle\phi_{3}|-\chi|\phi_{4}\rangle\langle\phi_{4}|$,
and we set $\chi\gg \Omega_{1}(t),\Omega_{2}(t),\lambda$ so that we
can perform the unitary transformation $U=e^{-iH_{0}t}$. By
discarding the terms with high oscillating frequency $\chi$, an
effective Hamiltonian is obtained
\begin{eqnarray}\label{eq3-7}
  H_{eff}&=&\frac{v}{\chi}|\phi_{0}\rangle(\Omega_{1}\langle\psi_{1}|+\Omega_{2}\langle\psi_{7}|)
           +\frac{1}{\sqrt{2}}|\mu_{1}\rangle(-\Omega_{1}\langle\psi_{1}|+\Omega_{2}\langle\psi_{7}|)  \cr\cr
         &&+\lambda|\mu_{1}\rangle\langle\mu_{2}|+H.c..
\end{eqnarray}
In fact, setting $\chi\gg \Omega_{1}(t),\Omega_{2}(t),\lambda$ is
helpful to shorten the operation time because when $v$ is too small,
traversing of photons between the cavities and the fiber will be
very difficult which increases the interaction time. Through
analyzing the proportions of the base vectors in Eq. (\ref{eq3-7})
in the eigenstates $|\Theta_{1}\rangle$ and $|\Theta_{2}\rangle$,
where $|\Theta_{1}\rangle$ and $|\Theta_{2}\rangle$ are the
eigenstates whose eigenvalues are closest to zero of $H_{eff}$, the
relation between the proportions of the states $|\phi_{0}\rangle$
and $|\mu_{1}\rangle$ is given:
\begin{eqnarray}\label{eq3-8}
  r=|\frac{P_{\phi_{0}}}{P_{\mu_{1}}}|=|\frac{\sqrt{6}
     [\sqrt{(5\Omega_{1}^{2}-5\Omega_{2}^{2})^{2}+4\Omega_{1}^{2}\Omega_{2}^{2}+12(\Omega_{1}^{2}+\Omega_{2}^{2})+36}+7]}{12(\Omega_{1}^2-\Omega_{2}^{2})}|.
\end{eqnarray}
We assume $v=g=1$ as a typical example here in order to simplify the
analysis. In general, to transfer the population from the initial
state $|\psi_{1}\rangle$ to the target state $|\psi_{7}\rangle$ via adiabatic passage, the
form of Rabi frequencies can be chosen as
$\Omega_{1}(t)=\Omega_{0}\sin{\beta}$ and
$\Omega_{2}(t)=\Omega_{0}\cos{\beta}$, where $\Omega_{0}$ denotes
the amplitude of the laser pulse and $\beta$ is a time-related
parameter. In this case,
\begin{eqnarray}\label{eq3-9}
  r&=&\frac{\sqrt{25\Omega_{0}^{4}\cos^{2}{2\beta}+\Omega_{0}^{4}\sin^{2}{2\beta}+12\Omega_{0}^{2}+36}+7}{2\sqrt{6}\Omega_{0}^{2}\cos{2\beta}} \cr\cr
   &=&\frac{\sqrt{24\Omega_{0}^{4}\cos^{2}{2\beta}+\Omega_{0}^{4}+12\Omega_{0}^{2}+36}+7}{2\sqrt{6}\Omega_{0}^{2}\cos{2\beta}}                   \cr\cr
   &\geq&\frac{\sqrt{25\Omega_{0}^{4}+12\Omega_{0}^{2}+36}+7}{2\sqrt{6}\Omega_{0}^{2}}.
\end{eqnarray}
It is obvious that when $r^{2}\gg 1$, the population of the state
$|\mu_{1}\rangle$ is far less than that of the state
$|\phi_{0}\rangle$. That means when $r^2\gg1$, the state
$|\mu_{1}\rangle$ is considered as negligible since the state
$|\phi_{0}\rangle$ is only limited populated during the evolution \cite{YHCYXQQCJSPRA14}.
Then the whole system can be divided into two parts which are
independent of each other: the main subsystem
$S_{m}=\{|\psi_{1}\rangle,|\psi_{7}\rangle,|\phi_{0}\rangle\}$ and
assistant subsystem $S_{a}=\{|\mu_{2}\rangle\}$. And the Hamiltonian
for the main subsystem is
\begin{eqnarray}\label{eq3-10}
  H_{m}=\frac{v}{\chi}|\phi_{0}\rangle(\Omega_{1}\langle\psi_{1}|+\Omega_{2}\langle\psi_{7}|)+H.c..
\end{eqnarray}
Then, as $H_{m}$ possesses SU(2) dynamical symmetry
\cite{YZLJQLHJWMKJGZPra96}, the invariant Hermitian operator $I(t)$
which satisfies $i\partial_{t}I(t)=[H_{m},I]$ can be easily given
\cite{XCETJGMPra11,YHCYXQQCJSPRA14}
\begin{eqnarray}\label{eq3-11}
  I(t)&=&\chi_{0}(\cos{\gamma}\sin{\beta}|\phi_{0}\rangle\langle\psi_{1}|+\cos{\gamma}\cos{\beta}|\phi_{0}\rangle\langle\psi_{7}|   \cr
       &&+i\sin{\gamma}|\psi_{7}\rangle\langle\psi_{1}|+H.c.),
\end{eqnarray}
where $\chi_{0}$ is an arbitrary constant with units of frequency to
keep $I(t)$ with dimensions of energy, $\gamma$ and $\beta$ are both
time-dependent auxiliary parameters. Through solving the relation
$i\partial_{t}I(t)=[H_{m},I]$, $\Omega_{1}$ and $\Omega_{2}$ are
obtained,
\begin{eqnarray}\label{eq3-12}
  \Omega_{1}(t)&=&\frac{\chi}{v}(\dot{\beta}\cot{\gamma}\sin{\beta}+\dot{\gamma}\cos{\beta}), \cr
  \Omega_{2}(t)&=&\frac{\chi}{v}(\dot{\beta}\cot{\gamma}\cos{\beta}-\dot{\gamma}\sin{\beta}),
\end{eqnarray}
where the dot represents a time derivative. The general solution of
the Schr\"{o}dinger equation
$i\partial_{t}|\psi\rangle=H_{m}|\psi\rangle$ with respect to the
instantaneous eigenstates of $I(t)$ is written as
\begin{eqnarray}\label{eq3-13}
  |\psi(t)\rangle=\sum_{n=0,\pm}C_{n}e^{i\alpha_{n}}|\tilde{\theta}_{n}(t)\rangle,
\end{eqnarray}
where $\alpha_{n}$ are the LR phases mentioned in Sec. \textrm{II}
and $|\tilde{\theta}_{n}\rangle$ are the instantaneous eigenstates
of $I(t)$
\begin{eqnarray}\label{eq3-14}
  |\tilde{\theta}_{0}\rangle&=&\cos{\gamma}\cos{\beta}|\psi_{1}\rangle-i\sin{\gamma}|\phi_{0}\rangle-\cos{\gamma}\sin{\beta}|\psi_{7}\rangle,  \cr\cr
  |\tilde{\theta}_{\pm}\rangle&=&\frac{1}{\sqrt{2}}[
                                                  (\sin{\gamma\cos{\beta}\pm i\sin{\beta}})|\psi_{1}\rangle
                                                 +i\cos{\gamma}|\phi_{0}\rangle                                                                \cr\cr
                                                 &&-(\sin{\gamma}\sin{\beta}\mp i\cos{\beta})|\psi_{7}\rangle
                                                  ].
\end{eqnarray}
To transfer population from the initial state $|\psi_{1}\rangle$ to
the target state $-|\psi_{7}\rangle$, a simple choice for the
parameters is
\begin{eqnarray}\label{eq3-15}
  \gamma=\epsilon,\ \beta=\pi t/2t_{f},
\end{eqnarray}
where $\epsilon$ is a time-independent small value and $t_{f}$ is
the interaction time. And we obtain
\begin{eqnarray}\label{eq3-16}
  \Omega_{1}(t)&=&\frac{\chi\pi\cot{\epsilon}}{2vt_{f}}\sin{\frac{\pi t}{2t_{f}}}, \cr\cr
  \Omega_{2}(t)&=&\frac{\chi\pi\cot{\epsilon}}{2vt_{f}}\cos{\frac{\pi t}{2t_{f}}}.
\end{eqnarray}
In the present case, when $t=t_{f}$,
\begin{eqnarray}\label{eq3-17}
  |\psi(t_{f})\rangle=
      \left(
     \begin{array}{c}
       i\sin{\epsilon}\sin{\alpha}                                                       \\
       -i\sin{\epsilon}\cos{\epsilon}+i\sin{\epsilon}\cos{\epsilon}\cos{\alpha}          \\
       -\cos^{2}{\epsilon}-\sin^{2}{\epsilon}\cos{\alpha}                                    \\
     \end{array}
    \right),
\end{eqnarray}
where $\alpha=\pi/(2\sin{\epsilon})=|\alpha_{\pm}|$. Hence, when we
choose $\alpha=2N\pi$ ($N=\pm1,\pm2\cdots$),
$|\psi(t_{f})\rangle=[0,0,-1]'=-|\psi_{7}\rangle$. Meanwhile, in the
assistant subsystem $S_{a}$, the time-dependence population of the
state $|\mu_{2}\rangle$ is mainly dominated by the dark state
evolution \cite{YHCYXQQCJSPRA14}, and when $t=t_{f}$, its population
becomes zero and the dark state also evolves into the state
$-|\psi_{7}\rangle$. That is, with joint efforts of both the
subsystems, the whole system fast evolves from the initial state
$|\psi_{1}\rangle$ to the final state $-|\psi_{7}\rangle$.

\section{Fast entangled states preparation in two separated cavities which are connected by a fiber}

\subsection{Bell states}
In this section, we will put forward two methods to generate Bell
states via the STAP proposed in Sec. III. The first method is
simple and easy. We only have to introduce an auxiliary ground state
$|a\rangle$ not interacting with other states in the atom 1 as shown
in Fig. \ref{model} (c) and set the initial state as follows
\begin{eqnarray}\label{eq4a-1}
  |\psi_{0}\rangle&=&\frac{1}{\sqrt{2}}(|f\rangle+|a\rangle)_{1}|g\rangle_{2}|0\rangle_{c_{1}}|0\rangle_{c_{2}}|0\rangle_{f}\cr\cr
                  &=&\frac{1}{\sqrt{2}}(|\psi_{1}\rangle+|\psi_{a}\rangle),
\end{eqnarray}
where
$|\psi_{a}\rangle=|a\rangle_{1}|g\rangle_{2}|0\rangle_{c_{1}}|0\rangle_{c_{2}}|0\rangle_{f}$.
Similar to the FPT in Sec. \textrm{III}, the term
$|f\rangle_{1}|g\rangle_{2}|0,0\rangle_{c_{1},c_{2}}|0\rangle_{f}$
will evolve along the STAP constructed above while the other term
$|a\rangle_{1}|g\rangle_{2}|0,0\rangle_{c_{1},c_{2}}|0\rangle_{f}$
will remain the same. The evolution of the system is governed by the
Hamiltonian in Eq. (\ref{eq3-1}). With the parameters we chosen,
when $t=t_{f}$, the final state of the system becomes
\begin{eqnarray}\label{eq4a-2}
  |\psi(t_{f})\rangle&=&\frac{1}{\sqrt{2}}[
                              (i\sin{\epsilon}\sin{\alpha})|\psi_{1}\rangle
                              -(i\sin{\epsilon}\cos{\epsilon}+i\sin{\epsilon}\cos{\epsilon}\cos{\alpha})|\phi_{0}\rangle    \cr\cr
                              &&-(\cos^{2}{\epsilon}-\sin^{2}{\epsilon}\cos{\alpha})|\psi_{7}\rangle
                              +|\psi_{a}\rangle].
\end{eqnarray}
When we choose $\epsilon=\arcsin{1/(4N)}$, Eq. (\ref{eq4a-2})
becomes the maximally entangled state
$|\psi(t_{f})\rangle=\frac{1}{\sqrt{2}}(-|\psi_{7}\rangle+|\psi_{a}\rangle)=|Bell\rangle$.

As the second method for generating a two-atom maximally entangled state,
we trap two atoms in the cavity $c_{2}$ while others are unchanged (the set-up diagram and the atomic level configuration for each atom are
similar to that in Sec. III).
In this case, the Hamiltonian in the interaction picture
for the whole system is
\begin{eqnarray}\label{eq4a-3}
  H_{i}&=&H_{al}+H_{ac}+H_{cf},                                                                                       \cr\cr
  H_{al}&=&\sum_{n=1}^{3}\Omega_{n}(t)|e\rangle_{n}\langle f|+H.c.,              \cr\cr
  H_{ac}&=&\lambda_{1}a_{1}|e\rangle_{1}\langle g|+\sum_{n=2}^{3}\lambda_{n}a_{2}|e\rangle_{n}\langle g|+H.c.,        \cr\cr
  H_{cf}&=&vb^{\dag}(a_{1}+a_{2})+H.c.,
\end{eqnarray}
and the closed subspace is spanned by
\begin{eqnarray}\label{eq4a-4}
  |\zeta_{1}\rangle&=&|f\rangle_{1}|g\rangle_{2}|g\rangle_{3}|0\rangle_{c_{1}}|0\rangle_{c_{2}}|0\rangle_{f}, \cr
  |\zeta_{2}\rangle&=&|e\rangle_{1}|g\rangle_{2}|g\rangle_{3}|0\rangle_{c_{1}}|0\rangle_{c_{2}}|0\rangle_{f}, \cr
  |\zeta_{3}\rangle&=&|g\rangle_{1}|g\rangle_{2}|g\rangle_{3}|1\rangle_{c_{1}}|0\rangle_{c_{2}}|0\rangle_{f}, \cr
  |\zeta_{4}\rangle&=&|g\rangle_{1}|g\rangle_{2}|g\rangle_{3}|0\rangle_{c_{1}}|0\rangle_{c_{2}}|1\rangle_{f}, \cr
  |\zeta_{5}\rangle&=&|g\rangle_{1}|g\rangle_{2}|g\rangle_{3}|0\rangle_{c_{1}}|1\rangle_{c_{2}}|0\rangle_{f}, \cr
  |\zeta_{6}\rangle&=&|g\rangle_{1}|e\rangle_{2}|g\rangle_{3}|0\rangle_{c_{1}}|0\rangle_{c_{2}}|0\rangle_{f}, \cr
  |\zeta_{7}\rangle&=&|g\rangle_{1}|f\rangle_{2}|g\rangle_{3}|0\rangle_{c_{1}}|0\rangle_{c_{2}}|0\rangle_{f}, \cr
  |\zeta_{8}\rangle&=&|g\rangle_{1}|g\rangle_{2}|e\rangle_{3}|0\rangle_{c_{1}}|0\rangle_{c_{2}}|0\rangle_{f}, \cr
  |\zeta_{9}\rangle&=&|g\rangle_{1}|g\rangle_{2}|f\rangle_{3}|0\rangle_{c_{1}}|0\rangle_{c_{2}}|0\rangle_{f}.
\end{eqnarray}
There are two eigenstates with null eigenvalues for the intermediate
Hamiltonian $H_{cf}+H_{ac}$ in this subspace. Choosing
$\lambda_{1}=\sqrt{2}\lambda_{2}=\sqrt{2}\lambda_{3}=\lambda$, we
orthogonalize these states and obtain a special dark state
$|S_{0}\rangle$ which will evolve into an independent subspace while
other states remain unchanged:
\begin{eqnarray}\label{eq4a-5}
  |S_{0}\rangle=\frac{v}{\sqrt{2v^2+\lambda^2}}(|\zeta_{2}\rangle
                         -\frac{\lambda}{v}|\zeta_{4}\rangle+\frac{1}{\sqrt{2}}|\zeta_{6}\rangle+\frac{1}{\sqrt{2}}|\zeta_{8}\rangle).
\end{eqnarray}
Introducing a vector
$|\varpi\rangle=(|\zeta_{6}\rangle+|\zeta_{8}\rangle)/\sqrt{2}$ and
eliminating the states which are irrelevant to the evolution,
eigenstates of the intermediate Hamiltonian which have the same form
with that in Eq. (\ref{eq3-3}) are obtained. Afterwards, setting
$\Omega_{2}(t)=\Omega_{3}(t)$, we can similarly rewrite the
Hamiltonian in Eq. (\ref{eq4a-1}) as
\begin{eqnarray}\label{eq4a-6}
  H_{re}^{T}&=&\frac{v}{\chi}|S_{0}\rangle(\Omega_{1}\langle\zeta_{1}|+\Omega_{2}\langle\varpi_{f}|)
         +\frac{1}{\sqrt{2}}|\mu_{1}\rangle(-\Omega_{1}\langle\zeta_{1}|+\Omega_{2}\langle\varpi_{f}|)               \cr\cr
          &&+\frac{\lambda}{\sqrt{2}\chi}|\mu_{3}\rangle(\Omega_{1}\langle\zeta_{1}|+\Omega_{2}\langle\varpi_{f}|)
          +\lambda|\tilde{\mu}_{1}\rangle\langle\tilde{\mu}_{2}|+\chi|\tilde{\mu}_{3}\rangle\langle\tilde{\mu}_{4}|+H.c.,
\end{eqnarray}
where
\begin{eqnarray}\label{eq4a-7}
  |\varpi_{f}\rangle&=&\frac{1}{\sqrt{2}}(|\zeta_{7}\rangle+|\zeta_{9}\rangle),                   \cr
  |\tilde{\mu}_{1}\rangle&=&\frac{1}{\sqrt{2}}(-|\zeta_{2}\rangle+|\varpi\rangle),                \cr
  |\tilde{\mu}_{2}\rangle&=&\frac{1}{\sqrt{2}}(-|\zeta_{3}\rangle+|\zeta_{5}\rangle)              \cr
  |\tilde{\mu}_{3}\rangle&=&\frac{\lambda}{\sqrt{4v^{2}+2\lambda^{2}}}(|\zeta_{2}\rangle
                            +\frac{2v}{\lambda}|\zeta_{4}\rangle+|\varpi\rangle),                 \cr
  |\tilde{\mu}_{4}\rangle&=&\frac{1}{\sqrt{2}}(|\zeta_{3}\rangle+|\zeta_{5}\rangle).
\end{eqnarray}
It is obvious that Eq. (\ref{eq4a-6}) equals to Eq. (\ref{eq3-4}).
The method used to construct the STAP in Sec. \textrm{III} also applies
to the present three-atom system. Therefore, with the Rabi
frequencies in Eq. (\ref{eq3-16}), when $t=t_{f}$, the system
evolves from the initial state $|\zeta_{1}\rangle$ to the final
state $|\varpi_{f}\rangle$. Meanwhile, the two atoms in the cavity
$c_{2}$ are turned into a maximally entangled state:
\begin{eqnarray}\label{eq4a-8}
  |Bell\rangle=\frac{1}{\sqrt{2}}(|f\rangle_{2}|g\rangle_{3}+|g\rangle_{2}|f\rangle_{3}).
\end{eqnarray}

\subsection{$M$-atom Greenberger-Horne-Zeilinger states}

Actually, the present STAP method in Sec. IV A can be generalized to
build an $M$-atom ($M \in \{1,2,3, \ldots , +\infty\}$) GHZ state.
We trap atom $1$ and atom $M$ in cavities $c_{1}$ and $c_{2}$,
respectively. The other $M-2$ atoms are discretionarily trapped in
the two cavities. The atomic level configuration is as the same as
we mentioned in the first STAP method for Bell-state generation. An
assistant ground state $|a\rangle$ is also required for each atom.
We assume atom $M$ is initially in the state $|g\rangle$ and the rest atoms are initially in an $(M-1)$-atom GHZ state
$\frac{1}{\sqrt{2}}(|a,g,g,\cdots,g\rangle+|f,a,a,\cdots,a\rangle)_{1,2,3,\cdots,M-1}$.
In this case, the term $|a,g,g,\cdots,g\rangle_{1,2,3,\cdots,M-1}|g\rangle_{M}$ will remain the same during the evolution.
And since the state $|a\rangle$ is only an assistant ground state, atoms $2,3,\cdots,M-1$ are actually irrelevant to the evolution,
only $|f,g\rangle_{1,M}$ participates in the evolution.
Therefore, the whole system can be
regarded as an effective two-atom system which is similar to the
first method for Bell-state generation. The shortcut is constructed
easily by setting the Rabi frequencies $\Omega_{1}(t)$ and
$\Omega_{M}(t)$ as the form of Eq. (\ref{eq3-16}). Similarly, when
$t=t_{f}$, the final state is obtained:
\begin{eqnarray}\label{eq4b-1}
  |\psi(t_{f})\rangle&=&\frac{1}{\sqrt{2}}[
                              (i\sin{\epsilon}\sin{\alpha})|\Psi_{1}\rangle
                              -(i\sin{\epsilon}\cos{\epsilon}+i\sin{\epsilon}\cos{\epsilon}\cos{\alpha})|\Phi_{0}\rangle    \cr\cr
                              &&-(\cos^{2}{\epsilon}-\sin^{2}{\epsilon}\cos{\alpha})|\Psi_{f}\rangle
                              +|\Psi_{a}\rangle].
\end{eqnarray}
where
\begin{eqnarray}\label{eq4b-2}
  |\Psi_{M}\rangle&=&|a,a,\cdots,a\rangle_{2,3,\cdots,M-1},                                                                                           \cr
  |\Psi_{1}\rangle&=&|f,g\rangle_{1,M}|\Psi_{M}\rangle|0,0\rangle_{c_{1},c_{2}}|0\rangle_{f},                                                         \cr
  |\Psi_{f}\rangle&=&|g,f\rangle_{1,M}|\Psi_{M}\rangle|0,0\rangle_{c_{1},c_{2}}|0\rangle_{f},                                                         \cr
  |\Psi_{a}\rangle&=&|a,g,g,\cdots,g\rangle_{1,2,3,\cdots,M}|0,0\rangle_{c_{1},c_{2}}|0\rangle_{f},                                                   \cr
  |\Phi_{0}\rangle&=&\frac{v}{\sqrt{2v^{2}+\lambda^{2}}}(
                                                         |e,g\rangle_{1,M}|\Psi_{M}\rangle|0,0\rangle_{c_{1},c_{2}}|0\rangle_{f}
                                                         +|g,e\rangle_{1,M}|\Psi_{M}\rangle|0,0\rangle_{c_{1},c_{2}}|0\rangle_{f}                     \cr
                                                         &&-\frac{\lambda}{v}|g,g\rangle_{1,M}|\Psi_{M}\rangle|0,0\rangle_{c_{1},c_{2}}|1\rangle_{f}).
\end{eqnarray}
There is no doubt that by choosing $\epsilon=\arcsin{(1/4N)}$
($N=\pm 1,\pm 2,\cdots$), Eq. (\ref{eq4b-1}) becomes an $M$-atom GHZ
state:
\begin{eqnarray}\label{eq4b-3}
  |GHZ\rangle=\frac{1}{\sqrt{2}}(|a,g,g,\cdots,g,g\rangle+|g,a,a,\cdots,a,f\rangle)_{1,2,3,\cdots,M-1,M}.
\end{eqnarray}
If we switch off the classical field $\Omega_{M}$, the state
$|f\rangle_{M}$ also can be regarded as an assistant ground state
$|a\rangle_{M}$. An $M$-atom GHZ state is generated by using an
$(M-1)$-atom GHZ state. And the transition ($|f\rangle\leftrightarrow|g\rangle$) of the atom $1$
can be driven by a
classical laser field without question. Therefore, when we trap one
more $\Lambda$-type atom whose initial state is $|g\rangle$ in the
cavity $c_{2}$, the present $M$-atom GHZ state can be converted into
an $(M+1)$-atom GHZ state.

\subsection{$M$-atom $W$ states}

We will show how to generate $M$-atom $W$ entangled state via
STAP in this section. The model we used is nearly the same as that
in Sec. \textrm{III} except that we trap $M$ identical
$\Lambda$-type atoms in the cavity $c_{2}$. In a typical setup with
neutral atoms in the cavity at least several microns apart direct
interaction between the atoms in the cavity $c_{2}$ are negligible.
The Hamiltonian of the whole system in the interaction picture can
be written as
\begin{eqnarray}\label{eq4c-1}
  H_{i}&=&H_{al}+H_{ac}+H_{cf},                                                                                       \cr\cr
  H_{al}&=&\Omega_{p}(t)|e\rangle_{p}\langle f|+\sum_{n=1}^{M}\Omega_{n}(t)|e\rangle_{n}\langle f|+H.c.,              \cr\cr
  H_{ac}&=&\lambda_{p}a_{1}|e\rangle_{p}\langle g|+\sum_{n=1}^{M}\lambda_{n}a_{2}|e\rangle_{n}\langle g|+H.c.,        \cr\cr
  H_{cf}&=&vb^{\dag}(a_{1}+a_{2})+H.c.,
\end{eqnarray}
where subscripts $p$ denotes the atom which is trapped in
the cavity $c_{1}$ and $n$ ($n \in [1,M]$) denotes the $n$th atom which
is trapped in the cavity $c_{2}$. For simplicity, we assume
$\Omega_{n}(t)=\Omega_{s}(t)$ and $\lambda_{n}=\lambda_{s}$. Since
the atoms in the cavity $c_{2}$ are indistinguishable, each atom
could be excited by the photon which is emitted from the atom
trapped in the cavity $c_{1}$ with the same probability; we can describe
the excited state of the atoms in the cavity $c_{2}$ as
\begin{eqnarray}\label{eq4c-2}
|\Phi_{e}\rangle=\frac{1}{\sqrt{M}}(|e,g,\cdots,g\rangle+|g,e,\cdots,g\rangle+\cdots+|g,g,\cdots,e\rangle)_{1,2,\cdots,M}.
\end{eqnarray}
The transition $|\Phi_{e}\rangle\leftrightarrow|\Phi_{g}\rangle$ is
coupled to the cavity mode with an effective coupling strength
$\sqrt{M}\lambda_{s}$, where
$|\Phi_{g}\rangle=|g,g,\cdots,g\rangle_{1,2,\cdots,M}$. In case of $\Omega_{n}=\Omega_{s}$,
$|\Phi_{e}\rangle$ might be driven to
\begin{eqnarray}\label{eq4c-3}
  |\Phi_{w}\rangle=\frac{1}{\sqrt{M}}(|f,g,\cdots,g\rangle+|g,f,\cdots,g\rangle+\cdots+|g,g,\cdots,f\rangle)_{1,2,\cdots,M}.
\end{eqnarray}
Then we assume the initial state is $|\Phi_{0}\rangle=|f\rangle_{p}|\Phi_{g}\rangle|0,0\rangle_{c_{1},c_{2}}|0\rangle_{f}$,
the whole system evolves in the subspace spanned by
\begin{eqnarray}\label{eq4c-4}
  |\Phi_{1}\rangle&=&|f\rangle_{p}|\Phi_{g}\rangle|0,0\rangle_{c_{1},c_{2}}|0\rangle_{f}, \cr
  |\Phi_{2}\rangle&=&|e\rangle_{p}|\Phi_{g}\rangle|0,0\rangle_{c_{1},c_{2}}|0\rangle_{f}, \cr
  |\Phi_{3}\rangle&=&|g\rangle_{p}|\Phi_{g}\rangle|1,0\rangle_{c_{1},c_{2}}|0\rangle_{f}, \cr
  |\Phi_{4}\rangle&=&|g\rangle_{p}|\Phi_{g}\rangle|0,0\rangle_{c_{1},c_{2}}|1\rangle_{f}, \cr
  |\Phi_{5}\rangle&=&|g\rangle_{p}|\Phi_{g}\rangle|0,1\rangle_{c_{1},c_{2}}|0\rangle_{f}, \cr
  |\Phi_{6}\rangle&=&|g\rangle_{p}|\Phi_{e}\rangle|0,0\rangle_{c_{1},c_{2}}|0\rangle_{f}, \cr
  |\Phi_{7}\rangle&=&|g\rangle_{p}|\Phi_{w}\rangle|0,0\rangle_{c_{1},c_{2}}|0\rangle_{f},
\end{eqnarray}
In this subspace, the Hamiltonians $H_{al}$ and $H_{ac}$ can be
written as
\begin{eqnarray}\label{eq4c-5}
  H_{al}&=&\Omega_{p}(t)|\Phi_{2}\rangle\langle \Phi_{1}|+\Omega_{s}(t)|\Phi_{6}\rangle\langle\Phi_{7}|+H.c.,     \cr
  H_{ac}&=&\lambda_{p}|\Phi_{2}\rangle\langle\Phi_{3}|+\sqrt{M}\lambda_{s}|\Phi_{6}\rangle\langle \Phi_{5}|+H.c..
\end{eqnarray}
When we choose $\lambda_{p}=\sqrt{M}\lambda_{s}=\lambda$ and
\begin{eqnarray}\label{4c-6}
  \Omega_{p}&=&\Omega_{1}=\frac{\chi\pi\cot{\epsilon}}{2vt_{f}}\sin{\frac{\pi t}{2t_{f}}},  \cr\cr
  \Omega_{s}&=&\Omega_{2}=\frac{\chi\pi\cot{\epsilon}}{2vt_{f}}\cos{\frac{\pi t}{2t_{f}}},
\end{eqnarray}
the total Hamiltonian in the present system equals to the
Hamiltonian in Eq. (\ref{eq3-1}). The dark state for this system is
given by:
\begin{eqnarray}\label{eq4c-7}
  |D_{W}\rangle=\frac{1}{N_{W}}(\Omega_{s}|\Phi_{1}\rangle-\Omega_{p}|\Phi_{7}\rangle+\frac{\sqrt{2}\Omega_{p}\Omega_{s}}{\lambda}|\mu_{2}\rangle),
\end{eqnarray}
where
$|\mu_{2}\rangle=(-|\Phi_{3}\rangle+|\Phi_{5}\rangle)/\sqrt{2}$ is
the intermediate state. And we have to make it independent from the
whole system by setting limiting condition which is similar to Eq. (\ref{eq3-8}).
Then the shortcuts are easily constructed as the same as what we do in Sec. \textrm{III}, and the evolution of the system is
approximatively described as
\begin{eqnarray}\label{eq4c-8}
  |\Phi(t)\rangle=\sum_{n=0,\pm}C_{n}e^{i\alpha_{n}}|\tilde{\theta}_{n}(t)\rangle.
\end{eqnarray}
When $t=t_{f}$, the final state becomes
$|\Phi(t_{f})\rangle=-|\Phi_{7}\rangle$, meanwhile, the $M$ atoms in the
cavity $c_{2}$ collapse to the $M$-qubit $W$ state
\begin{eqnarray}\label{eq4c-9}
  |W\rangle&=&\frac{1}{\sqrt{M}}(|f,g,g,\cdots,g\rangle+|g,f,g,\cdots,g\rangle+|g,g,f,\cdots,g\rangle   \cr
            &&+\cdots+|g,g,g,\cdots,f\rangle)_{1,2,3,\cdots,M}.
\end{eqnarray}

\section{Fast transition of two-atom entangled state via shortcuts to adiabatic passage}

In the following, we discuss the fast transition of the entangled
state from the cavity $c_{1}$ to $c_{2}$ via STAP. In this
section, we assume that there are four identical atoms, where two of
them are trapped in the cavity $c_{1}$ and the other two are trapped
in the cavity $c_{2}$. The atomic level configuration is the same as
that in Sec. \textrm{III}. Hence, the Hamiltonian of the present
system in the interaction picture is
\begin{eqnarray}\label{eq5-1}
  H_{i}&=&H_{al}+H_{ac}+H_{cf},                                  \cr\cr
  H_{al}&=&\sum_{k=1}^{4}{\Omega_{ak}|e\rangle_{k}\langle f|}+H.c., \cr\cr
  H_{ac}&=&\sum_{k=1,2}{\lambda_{k} a_{1}|e\rangle_{k}\langle g|}
          +\sum_{k=3,4}{\lambda_{k} a_{2}|e\rangle_{k}\langle g|}+H.c.,  \cr\cr
  H_{cf}&=&vb^{\dag}(a_{1}+a_{2}).
\end{eqnarray}
We assume that the initial state is
$|\Psi_{0}\rangle=\frac{1}{\sqrt{2}}(|fg\rangle+|gf\rangle)_{1,2}|gg\rangle_{3,4}|0,0\rangle_{c_{1},c_{2}}|0\rangle_{f}$
which is a two-atom entangled state. The system evolves into the
subspace spanned by
\begin{eqnarray}\label{eq5-2}
  |\varphi_{1}\rangle&=&|fg\rangle_{1,2}|gg\rangle_{3,4}|0,0\rangle_{c_{1},c_{2}}|0\rangle_{f},  \cr
  |\varphi_{2}\rangle&=&|eg\rangle_{1,2}|gg\rangle_{3,4}|0,0\rangle_{c_{1},c_{2}}|0\rangle_{f},  \cr
  |\varphi_{3}\rangle&=&|gg\rangle_{1,2}|gg\rangle_{3,4}|1,0\rangle_{c_{1},c_{2}}|0\rangle_{f},  \cr
  |\varphi_{4}\rangle&=&|ge\rangle_{1,2}|gg\rangle_{3,4}|0,0\rangle_{c_{1},c_{2}}|0\rangle_{f},  \cr
  |\varphi_{5}\rangle&=&|gf\rangle_{1,2}|gg\rangle_{3,4}|0,0\rangle_{c_{1},c_{2}}|0\rangle_{f},  \cr
  |\varphi_{6}\rangle&=&|gg\rangle_{1,2}|gg\rangle_{3,4}|0,0\rangle_{c_{1},c_{2}}|1\rangle_{f},  \cr
  |\varphi_{7}\rangle&=&|gg\rangle_{1,2}|eg\rangle_{3,4}|0,0\rangle_{c_{1},c_{2}}|0\rangle_{f},  \cr
  |\varphi_{8}\rangle&=&|gg\rangle_{1,2}|fg\rangle_{3,4}|0,0\rangle_{c_{1},c_{2}}|0\rangle_{f},  \cr
  |\varphi_{9}\rangle&=&|gg\rangle_{1,2}|gg\rangle_{3,4}|0,1\rangle_{c_{1},c_{2}}|0\rangle_{f},  \cr
  |\varphi_{10}\rangle&=&|gg\rangle_{1,2}|ge\rangle_{3,4}|0,0\rangle_{c_{1},c_{2}}|0\rangle_{f}, \cr
  |\varphi_{11}\rangle&=&|gg\rangle_{1,2}|gf\rangle_{3,4}|0,0\rangle_{c_{1},c_{2}}|0\rangle_{f}.
\end{eqnarray}
We regard the atoms in the same cavity as an integral whole as they
are indistinguishable. Like Sec. \textrm{IV}, we describe the
excited states of the atoms in the cavities as
\begin{eqnarray}\label{eq5-3}
  |\Psi_{2}\rangle=\frac{1}{\sqrt{2}}(|eg\rangle+|ge\rangle)_{1,2}|gg\rangle_{3,4},   \cr
  |\Psi_{4}\rangle=\frac{1}{\sqrt{2}}(|eg\rangle+|ge\rangle)_{3,4}|gg\rangle_{1,2}.
\end{eqnarray}
The atomic transitions
$|\Psi_{2}\rangle\leftrightarrow|\Psi_{g}\rangle$ is coupled to the
cavity mode with an effective coupling strength $\lambda$, so is
$|\Psi_{4}\rangle\leftrightarrow|\Psi_{g}\rangle$, where
$|\Psi_{g}\rangle=|gggg\rangle_{1,2,3,4}$ and
$\lambda_{k}=\lambda/\sqrt{2}$. By choosing
$\Omega_{a1,a2}=\Omega_{p}$ and $\Omega_{a3,a4}=\Omega_{s}$, we
rewrite the Hamiltonian in Eq. (\ref{eq5-1}) with the basis vectors
in Eqs. (\ref{eq5-2}) and (\ref{eq5-3}) as
\begin{eqnarray}\label{eq5-4}
  H_{i}&=&H_{al}+H_{ac}+H_{cf},                                                                                                              \cr\cr
  H_{al}&=&\Omega_{p}|\Psi_{2}\rangle|cf_{0}\rangle\langle\Psi_{0}|+\Omega_{s}|\Psi_{4}\rangle|cf_{0}\rangle\langle\Psi_{f}|+H.c.,           \cr\cr
  H_{ac}&=&\lambda|\Psi_{2}\rangle|cf_{0}\rangle\langle\varphi_{3}|+\lambda|\Psi_{4}\rangle|cf_{0}\rangle\langle\varphi_{9}|+H.c., \cr\cr
  H_{cf}&=&vb^{\dag}(a_{1}+a_{2}),
\end{eqnarray}
where
$|\Psi_{f}\rangle=\frac{1}{\sqrt{2}}|gg\rangle_{1,2}(|eg\rangle+|ge\rangle)_{3,4}|00\rangle_{c_{1},c_{2}}|0\rangle_{f}$
and $|cf_{0}\rangle=|00\rangle_{c_{1},c_{2}}|0\rangle_{f}$. It is
obvious that the present Hamiltonian has the same form with that in
Eq. (\ref{eq3-1}). When we choose
\begin{eqnarray}\label{eq5-5}
  \Omega_{p}&=&\Omega_{1}=\frac{\chi\pi\cot{\epsilon}}{2vt_{f}}\sin{\frac{\pi t}{2t_{f}}},  \cr\cr
  \Omega_{s}&=&\Omega_{2}=\frac{\chi\pi\cot{\epsilon}}{2vt_{f}}\cos{\frac{\pi t}{2t_{f}}},
\end{eqnarray}
the shortcut is
obtained. The whole system is approximatively described as
\begin{eqnarray}\label{eq5-6}
  |\Psi(t)\rangle=\sum_{n=0,\pm}C_{n}e^{i\alpha_{n}}|\tilde{\theta}_{n}(t)\rangle.
\end{eqnarray}
When $t=t_{f}$, the finial state is
\begin{eqnarray}\label{eq5-7}
  |\Psi(t_{f})\rangle=|\Psi_{f}\rangle=\frac{1}{\sqrt{2}}|gg\rangle_{1,2}(|eg\rangle+|ge\rangle)_{3,4}|00\rangle_{c_{1},c_{2}}|0\rangle_{f}.
\end{eqnarray}
In fact, the present scheme can be generalized to implement the
transition of a $M$-atom $W$ entangled state via trapping $M$ atoms
in respective cavities $c_{1}$ and $c_{2}$, and setting
$\lambda_{k}=\lambda/\sqrt{M}$.

\section{Numerical simulation and discussion}

\begin{figure}
 \scalebox{0.19}{\includegraphics {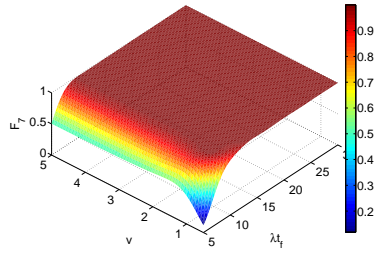}}
 \caption{The fidelity of the target state $F_{7}$ versus parameters $v$ and $\lambda t_{f}$.}
 \label{Fvgt}
\end{figure}

\begin{figure}
 \renewcommand\figurename{\small FIG.}
 \centering \vspace*{8pt} \setlength{\baselineskip}{10pt}
 \subfigure[]{
 \includegraphics[scale = 0.19]{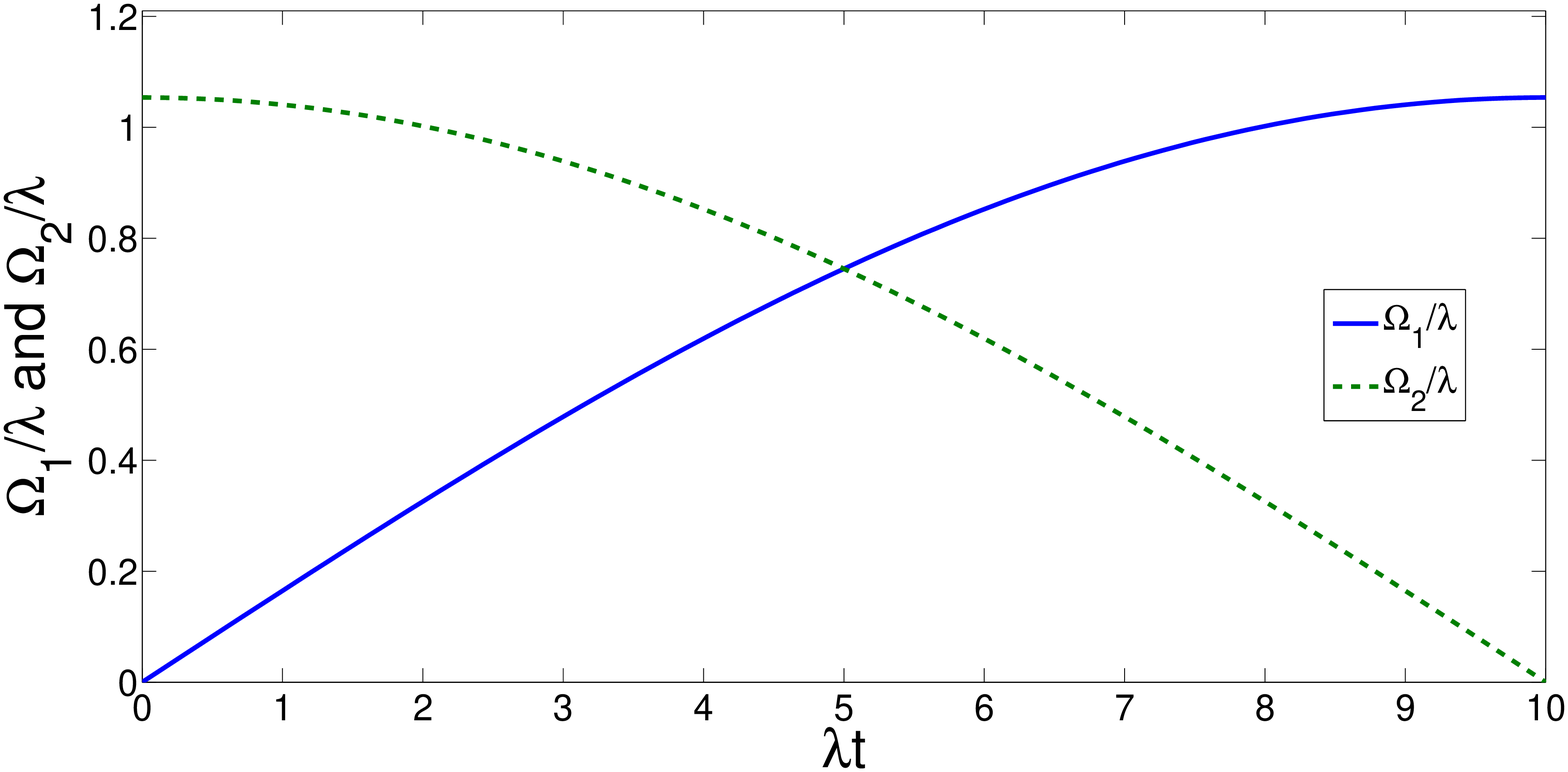}}
 \subfigure[]{
 \includegraphics[scale = 0.19]{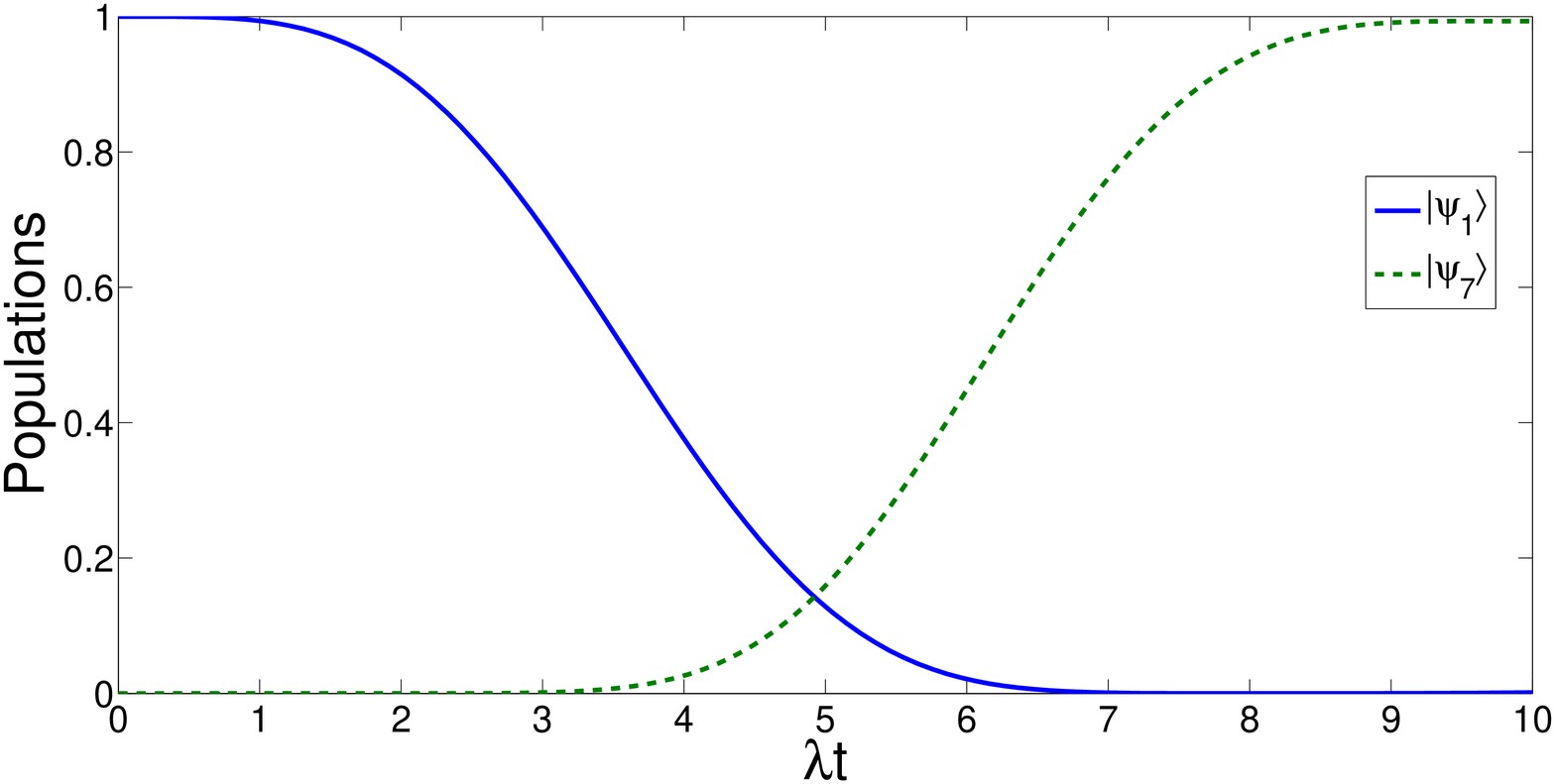}}
 \subfigure[]{
 \includegraphics[scale = 0.19]{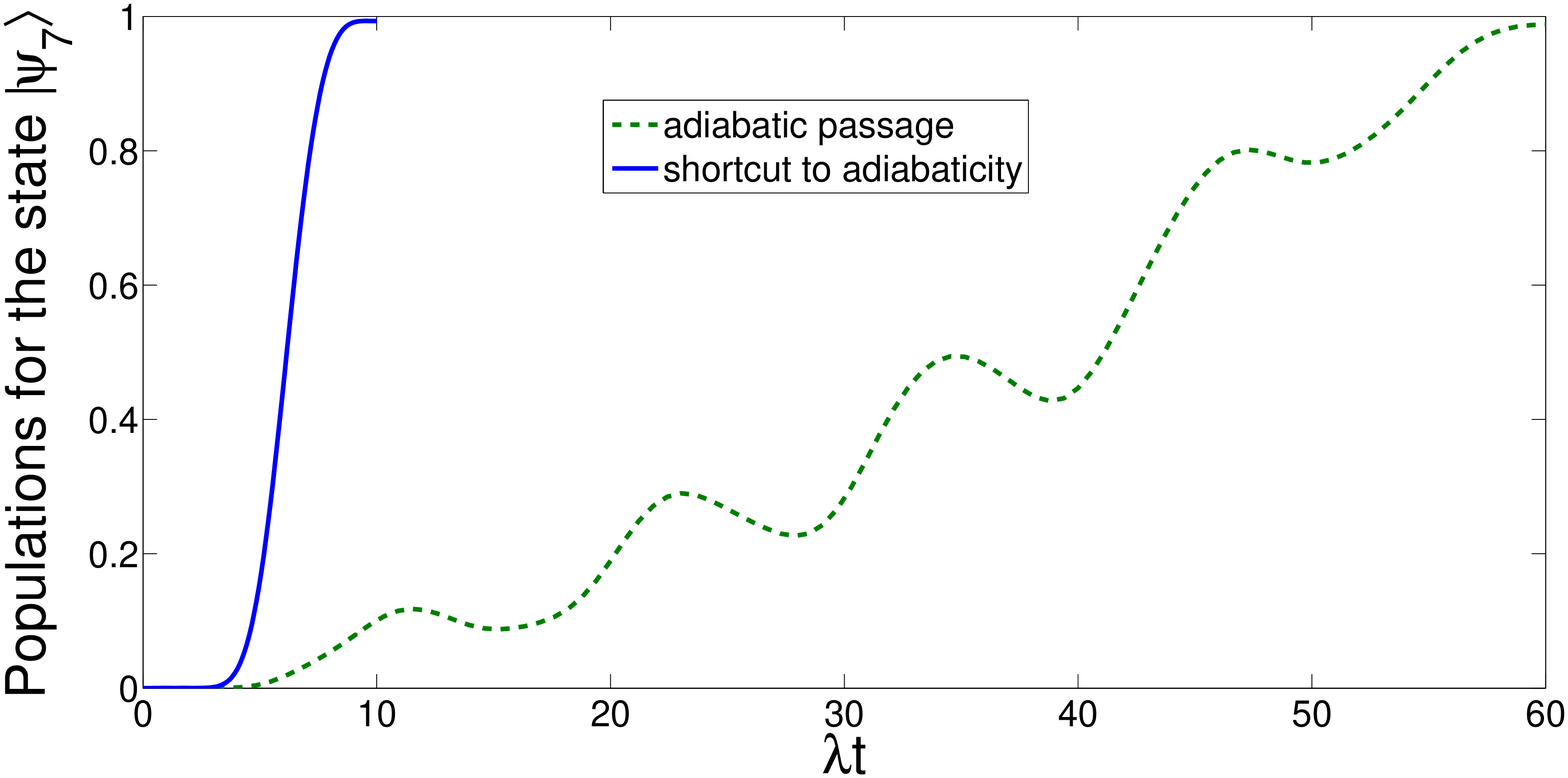}}
 \subfigure[]{
 \includegraphics[scale = 0.19]{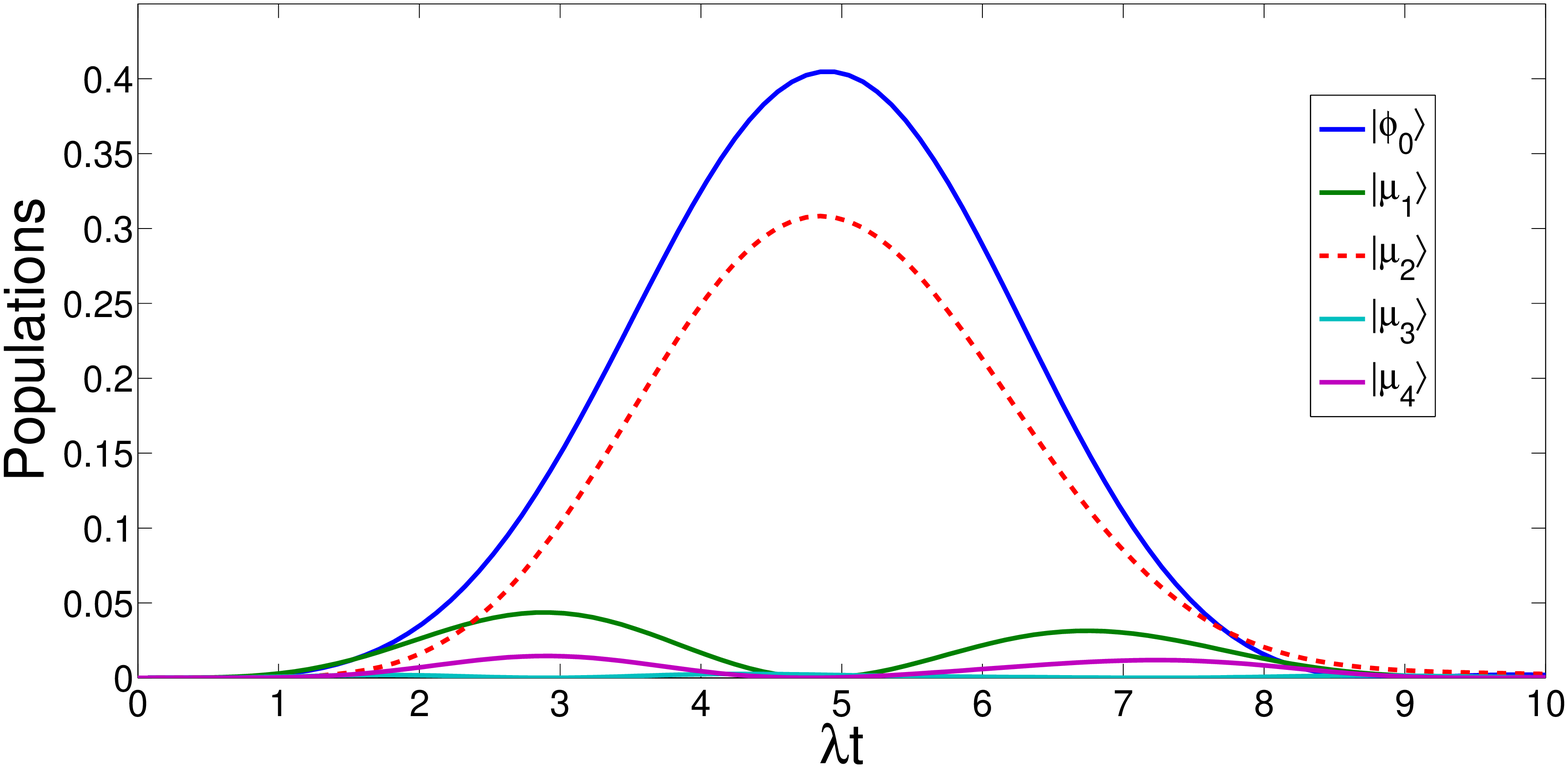}}
 \caption{
    (a) Dependence on $\lambda t$ of $\Omega_{1}(t)/\lambda$ and $\Omega_{2}(t)/\lambda$.
    (b) Dependence on $\lambda t$ of the populations for the initial state $|\psi_{1}\rangle$ and the target state $|\psi_{7}\rangle$.
    (c) The comparison of the operation times required for achieving the target state via adiabatic method with that via the present STAP method.
    (d) Dependence on $\lambda t$ of the populations for the intermediate states.
          }
\label{p0p7}
\end{figure}

We will first analyze the relation between the cavity-fiber coupling
$v$ and the interaction time $t_{f}$ as $v$ plays a great important
role in the evolution. The fidelity $F_{7}$ of the target state
$|\psi_{7}\rangle $  versus $v$ and $\lambda t_{f}$ is shown in Fig.
\ref{Fvgt} when $\epsilon=\arcsin{0.25}$, where the fidelity of a
state is given through the relation
$F=|\langle\psi|\rho(t_{f})|\psi\rangle|$. Figure 2 shows increasing
the value of $v$ does not help to shorten the interaction time,
which is different from what we mentioned above in Sec.
\textrm{III}. The reason is that the relation between the coupling
$v$ and the amplitude of the laser pulses $\Omega_{0}$ at that time
was not taken into consideration. Known from Refs.
\cite{XCILARDGJGMPrl10,AdCPra11,XCETJGMPra11,XCJGMPra12,MLYXLTSJSNBAPra14,YHCYXQQCJSPRA14},
shortening the time requires increasing the amplitude of the laser
pulses. The amplitude of the laser pulses in Eq. (\ref{eq3-16})
inverses proportion to the coupling $v$, and the smaller the
amplitude is, the longer the interaction time is. Consequently it is
wise to choose $v=g$ in our method. What is more, Fig. \ref{Fvgt}
shows that in the present case, the shortest interaction time
required for an ideal population transfer from $|\psi_{1}\rangle$ to
$|\psi_{7}\rangle$ is only about $8/\lambda$. The time dependences
of $\Omega_{1}(t)/\lambda$ and $\Omega_{2}(t)/\lambda$ are shown in
Fig. \ref{p0p7} (a) versus $\lambda t$ when
$\epsilon=\arcsin{0.25}$, $t_{f}=10/\lambda$ and $v=\lambda$. The
amplitude of the laser pulses $\Omega_{0}$ is $1.05\lambda$ which
meets the conditions mentioned above. And such an intensity is safe
to assume linear optic models. Similar laser amplitude has been used
and discussed in Refs.
\cite{YHCYXQQCJSPRA14,ADBABRMTENHJKPrl07,LBCMYYGWLQHDXMLPra07,JSYXXDSHSSPra12,JSXDSQXMLLZYXHSSPra13}.
In Ref. \cite{JSYXXDSHSSPra12} proposed by Song \emph{et al.}, even
high drive intensity $\Omega_{0}=\sqrt{5\sqrt{2}}\lambda\approx2.6591\lambda$ was
used. Fig. \ref{p0p7} (b) shows the time evolution of the
populations in states $|\psi_{1}\rangle$ and $|\psi_{7}\rangle$. We
contrast the interaction time required for achieving the target
state via adiabatic process with the present STAP method in Fig.
\ref{p0p7} (c). The result obviously shows that the present STAP
method effectively shorten the interaction time than the adiabatic
method even with the same laser intensity (we choose
$\Omega_{1}=\lambda\sin{\beta}$ and $\Omega_{2}=\lambda\cos{\beta}$
in the adiabatic method). A similar process in an adiabatic scheme
in Ref. \cite{LBCMYYGWLQHDXMLPra07} shows that the interaction time
required to achieve the target state is almost $t_{f}=250/\lambda$
when all the populations of the excited states should be restrained
to reduce the influence of dissipation. To demonstrate the
conditions for the STAP process are fulfilled, especially, the
condition that the state $|\mu_{1}\rangle$ is negligible during the
evolution, we plot Fig. \ref{p0p7} (d) in case of
\{$\epsilon=\arcsin{0.25},t_{f}=10/\lambda$\}. We find that the
states $|\mu_{3}\rangle$ and $|\mu_{4}\rangle$ remain negligible all
the time and the state $|\mu_{1}\rangle$ is very slightly populated
but still considered as negligible since the maximum value of its
population is only $4.36\%$. Besides, from Eq. (\ref{eq3-5}), it is
evident that the population of the state $|\mu_{2}\rangle$ has a
spacial relationship with $\lambda t_{f}$:
\begin{eqnarray}\label{eq6-1}
  P=|\frac{\sqrt{2}\Omega_{1}\Omega_{2}}{N_{D}\lambda}|^{2}=|\frac{2C^{2}}{(\lambda t_{f})^{2}+2C^{2}}|,
\end{eqnarray}
where $C=(\chi\pi\cot{\epsilon}\sin{\beta}\cos{\beta})/2v$. Though
the STAP process is constructed based on nonadiabatic processes, the
dark state $|D\rangle$ still can approximatively describe the
evolution of the system because the adiabatic condition has not been
absolutely destroyed. That is also why Eq. (\ref{eq6-1}) is still useful to
describe the population of the state $|\mu_{2}\rangle$. And the
result of Eq. (\ref{eq6-1}) shows that there is an inverse
relationship between $P$ and $\lambda t_{f}$, that means, increasing
the population of the state $|\mu_{2}\rangle$ is a simple and
effective way to shorten the interaction time. Moreover, we have
mentioned above that the state $|\mu_{2}\rangle$ can only be
transformed from the state $|\mu_{1}\rangle$. Therefore, under the
premise of the negligible state $|\mu_{1}\rangle$ during the
evolution, properly increasing the population of the state
$|\mu_{1}\rangle$ is of helpful to shorten the interaction time
(this result also applies to most of the adiabatic schemes when the
decoherence is not taken into consideration).

\begin{figure}
 \renewcommand\figurename{\small FIG.}
 \centering \vspace*{8pt} \setlength{\baselineskip}{10pt}
 \subfigure[]{
 \includegraphics[scale = 0.19]{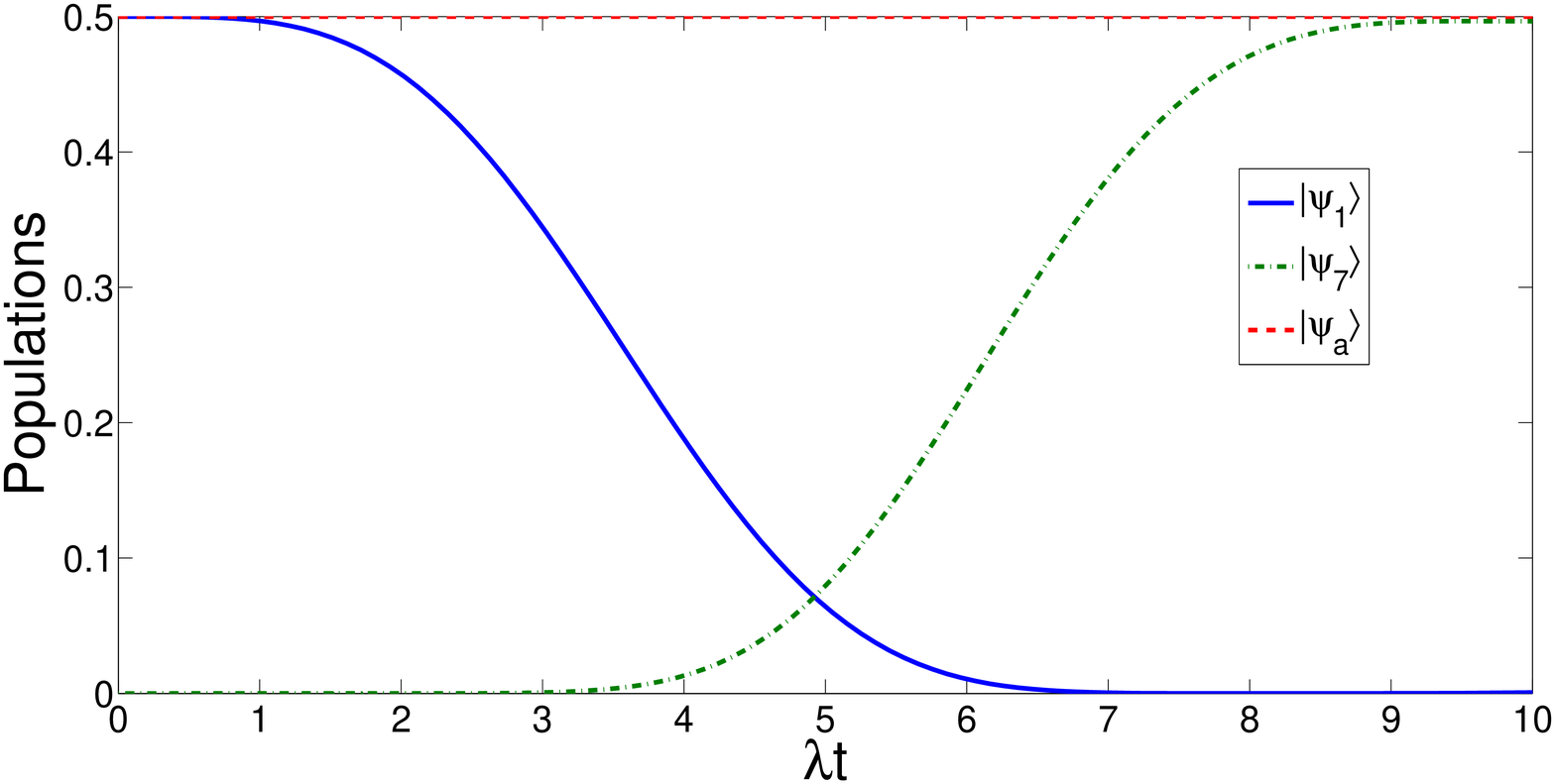}}
 \subfigure[]{
 \includegraphics[scale = 0.19]{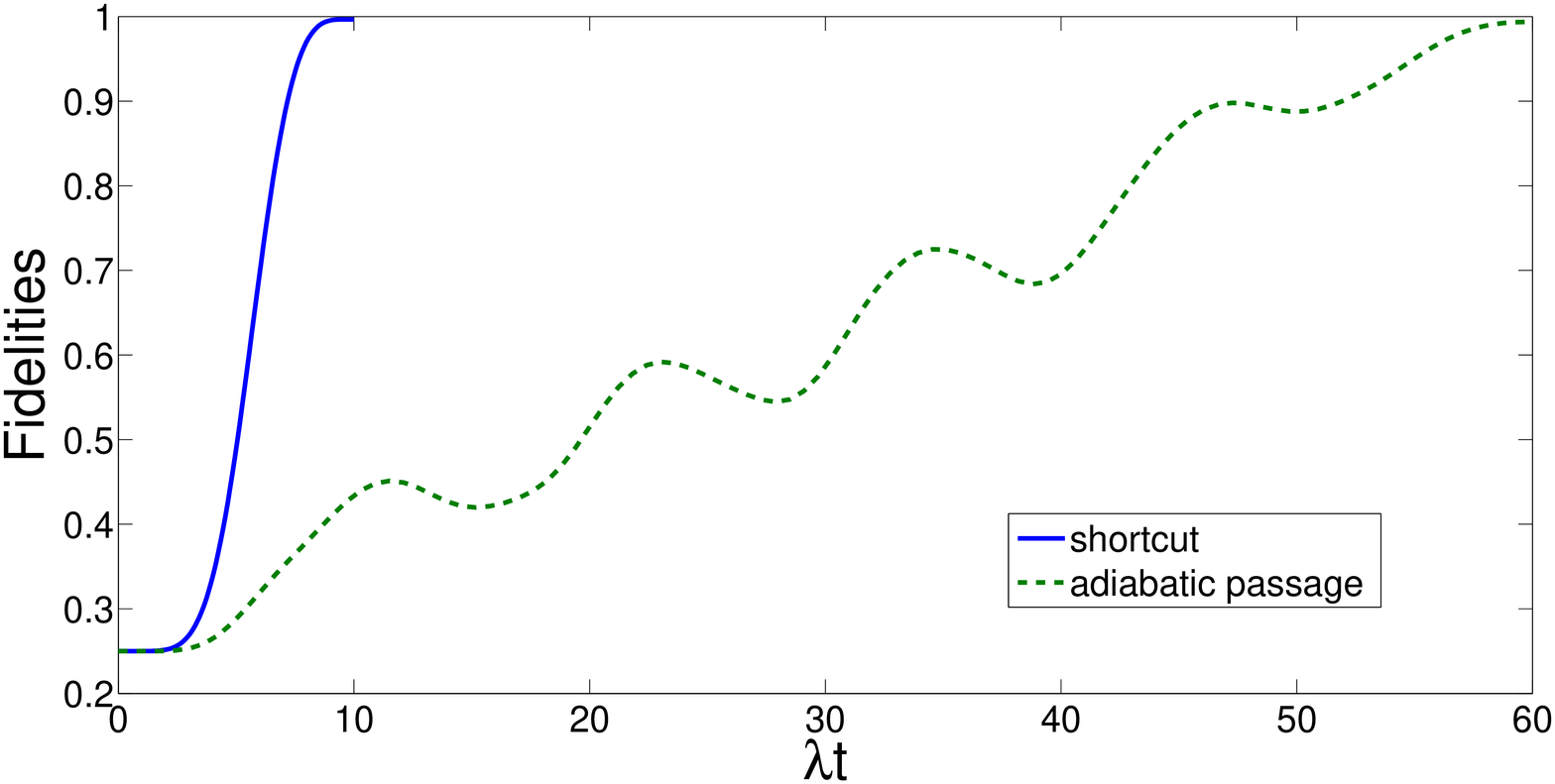}}
 \subfigure[]{
 \includegraphics[scale = 0.19]{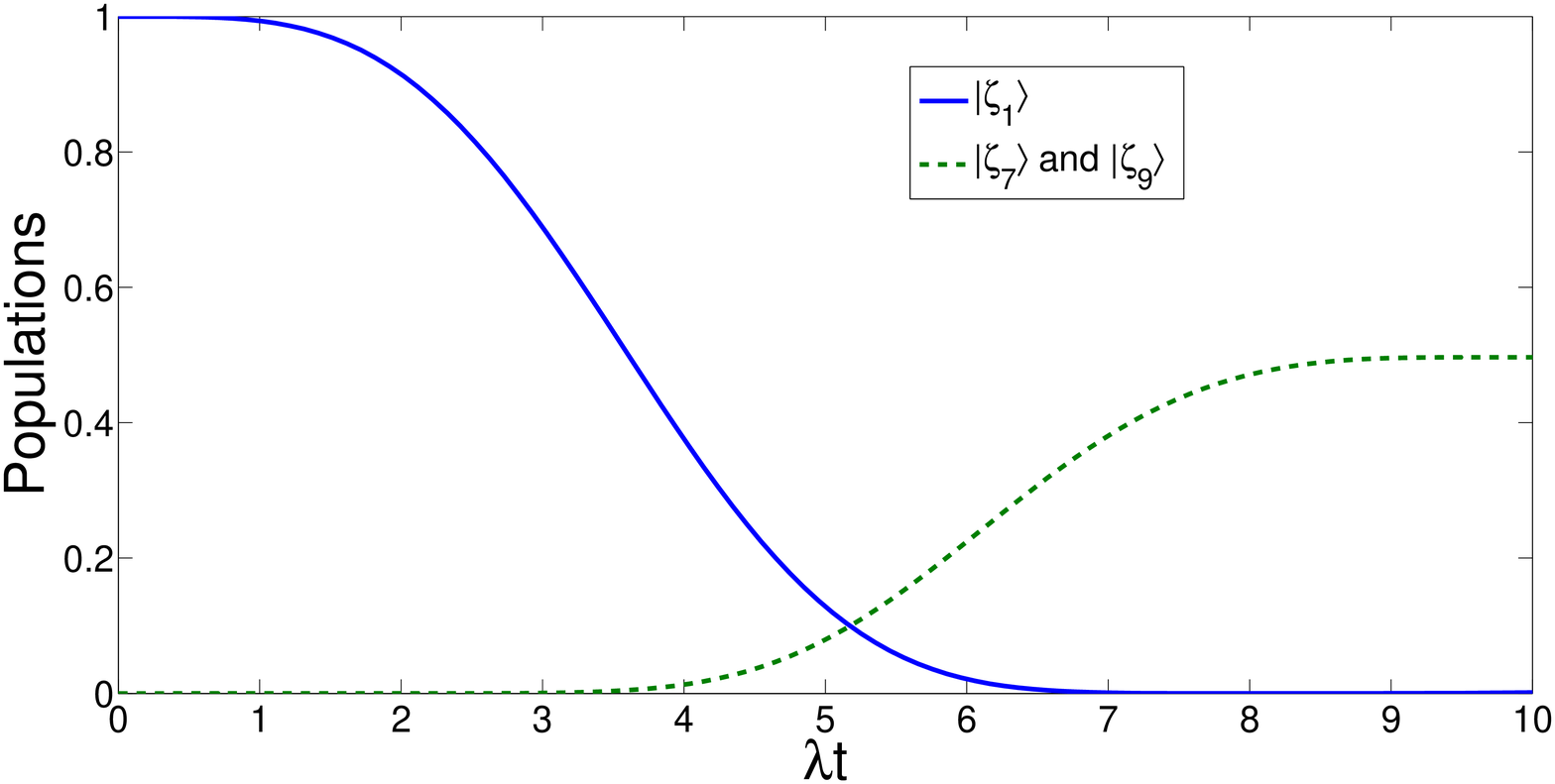}}
 \subfigure[]{
 \includegraphics[scale = 0.19]{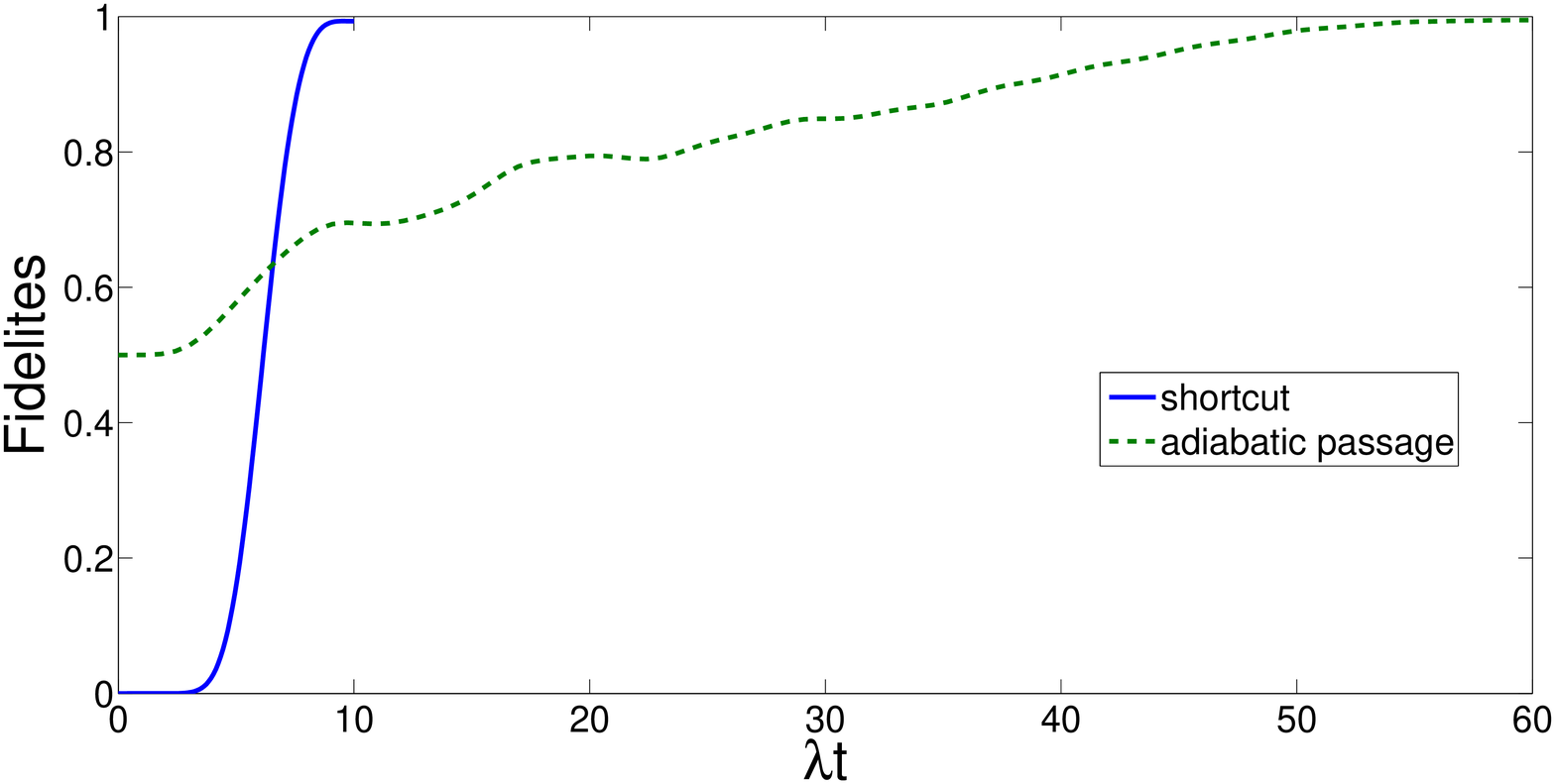}}
 \caption{
    (a) Time evolution of the populations for the states $|\phi_{1}\rangle$, $|\psi_{7}\rangle$, and $|\psi_{a}\rangle$ in the first scheme of Bell-state generation.
    (b) Time evolution of the fidelities of Bell stats via STAP and adiabatic methods.
    (c) Time evolution of the populations for the states $|\zeta_{1}\rangle$, $|\zeta_{7}\rangle$, and $|\zeta_{9}\rangle$ in the second scheme of Bell-state generation.
    (d) Time evolution of the fidelities of Bell stats via STAP and adiabatic methods.
          }
\label{Bell}
\end{figure}

In Fig. \ref{Bell}, we analyze the efficiency of the STAP for
the generation of Bell entangled states. Figure. \ref{Bell} (a)
displays the time evolution of the populations in states
$|\psi_{1}\rangle$, $|\psi_{7}\rangle$ and $|\psi_{a}\rangle$ in the
first STAP for Bell-state generation, and we give a comparison
between the performances of the fidelity of the two-atom Bell state
based on the first STAP method for Bell-state generation and an
adiabatic passage method governed by an Hamiltonian whose form is
the same as that in Eq. (\ref{eq3-1}) with parameters
$\{\Omega_{1}=\lambda\sin{\beta},\Omega_{2}=\lambda\cos{\beta},t_{f}=60/\lambda\}$
satisfying the adiabatic condition. Figure. \ref{Bell} (c) shows the
populations in states $|\zeta_{1}\rangle$, $|\zeta_{7}\rangle$ and
$|\zeta_{9}\rangle$ in the second STAP for Bell-state generation,
and the second STAP method is contrasted with adiabatic passage
method for the two-atom Bell state, demonstrating that the STAP
method is much faster when one more atom is used. The green dashed
curve in Fig. \ref{Bell} (d) is plotted based on a model with two
atoms trapped in a cavity whose interaction Hamiltonian is
$H=\sum_{k=1,2}\Omega_{k}|e\rangle_{k}\langle f|+\lambda
a|e\rangle_{k}\langle g|+H.c.$, and the parameters
$\{\Omega_{1}=0.5\lambda(\sin{\beta})^{1.2},\Omega_{2}=\lambda(\cos{\beta})^{1.2},t_{f}=60/\lambda\}$
are set to satisfy the adiabatic condition. These models and
parameters are chosen according to the principle of fair comparison.
Both Figs. \ref{Bell} (a) and (b), beyond doubt, give a result that
the STAP methods can generate Bell-entangled states much faster than
the adiabatic methods without requiring extra complex operations and
harsh conditions.

\begin{figure}
 \renewcommand\figurename{\small FIG.}
 \centering \vspace*{8pt} \setlength{\baselineskip}{10pt}
 \subfigure[]{
 \includegraphics[scale = 0.19]{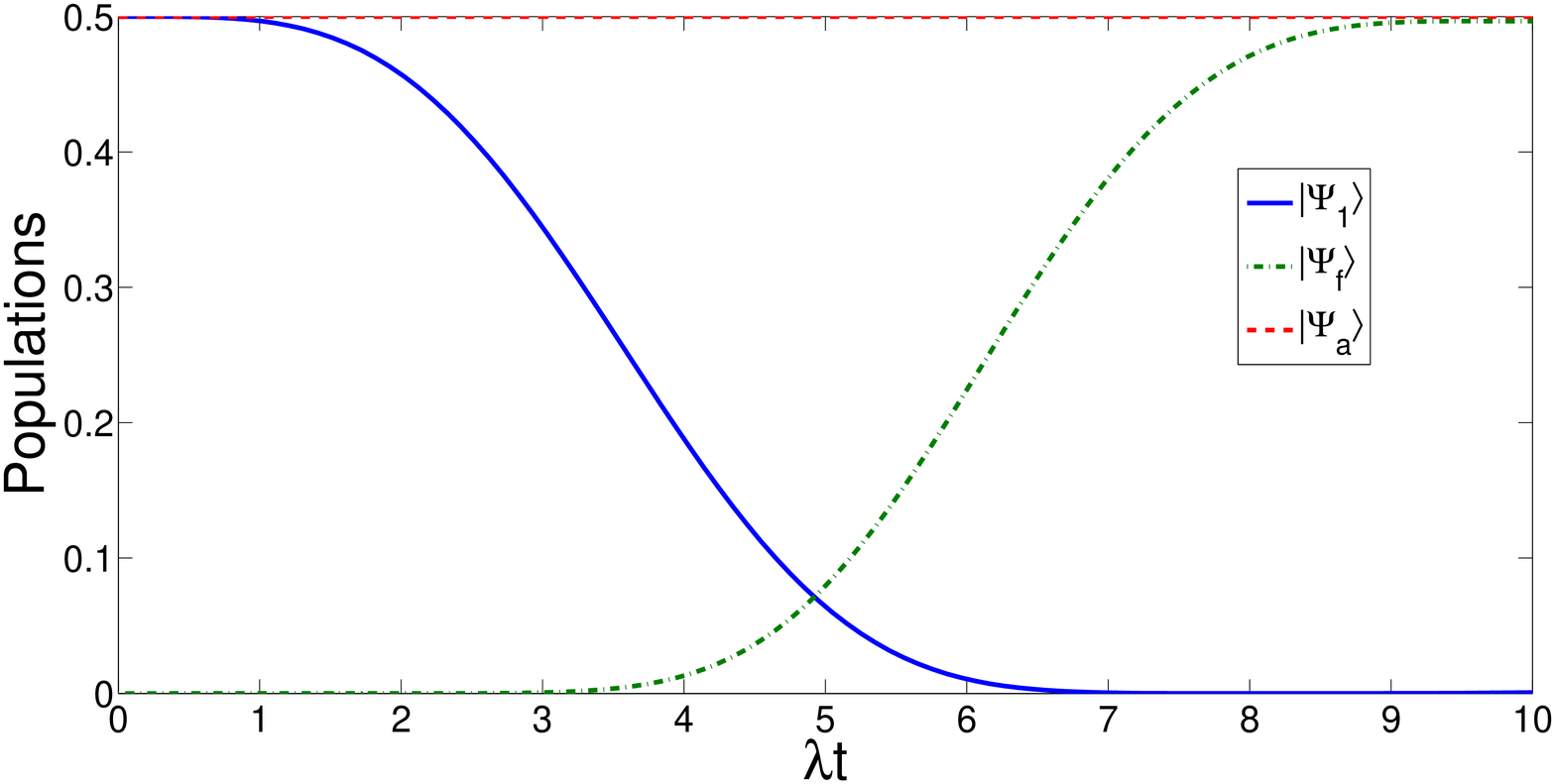}}
 \subfigure[]{
 \includegraphics[scale = 0.19]{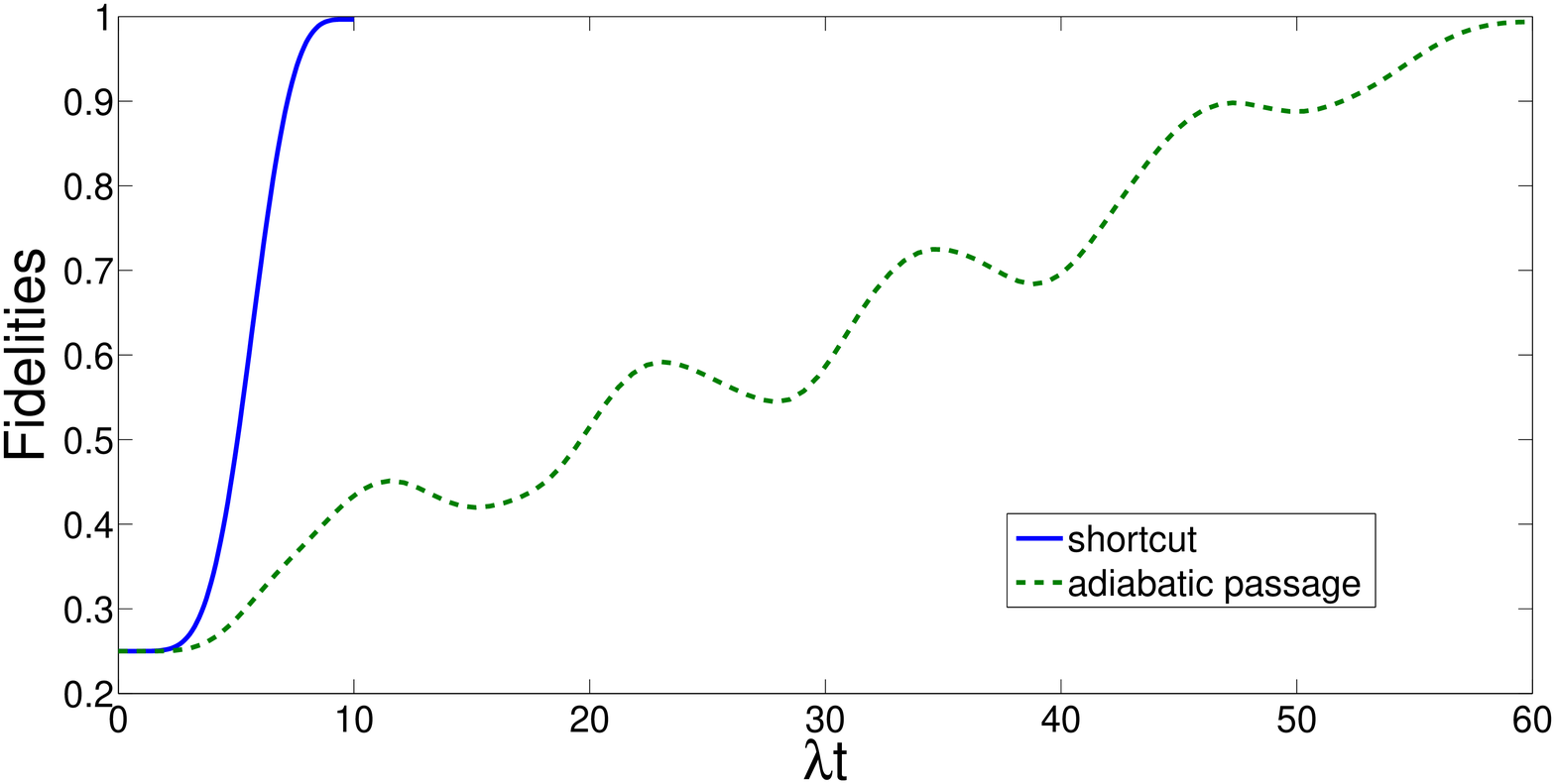}}
 \subfigure[]{
 \includegraphics[scale = 0.19]{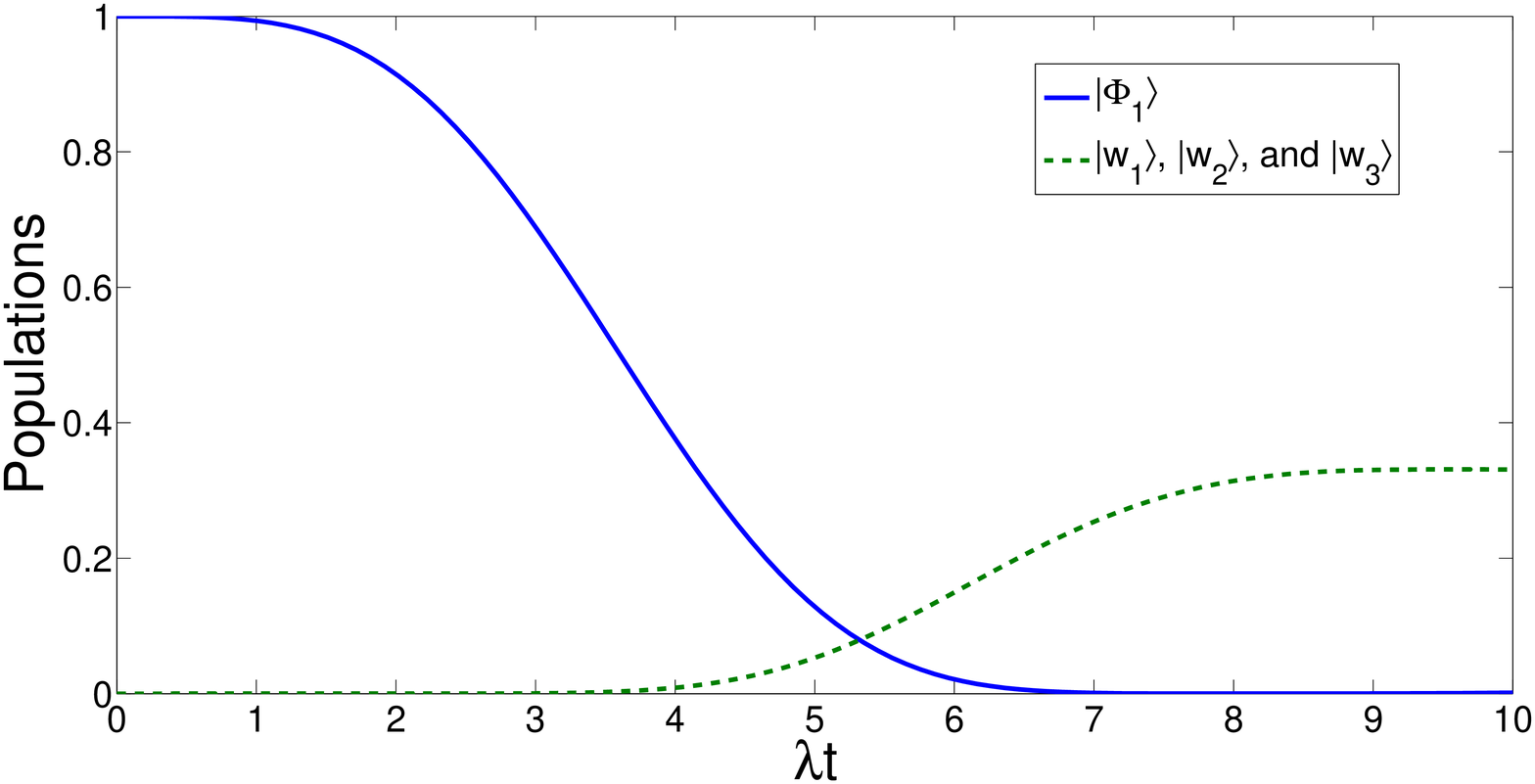}}
 \subfigure[]{
 \includegraphics[scale = 0.19]{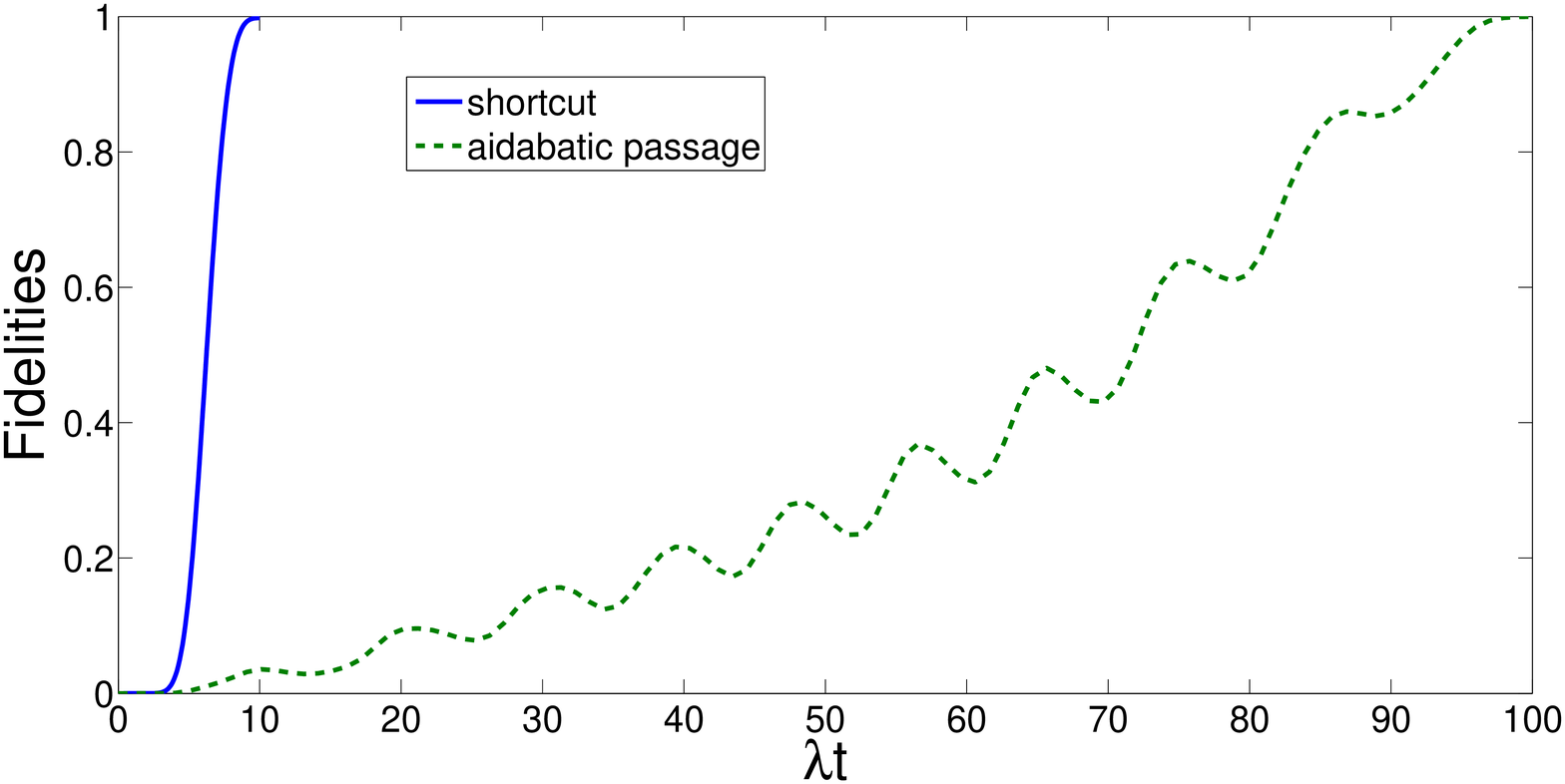}}
 \subfigure[]{
 \includegraphics[scale = 0.19]{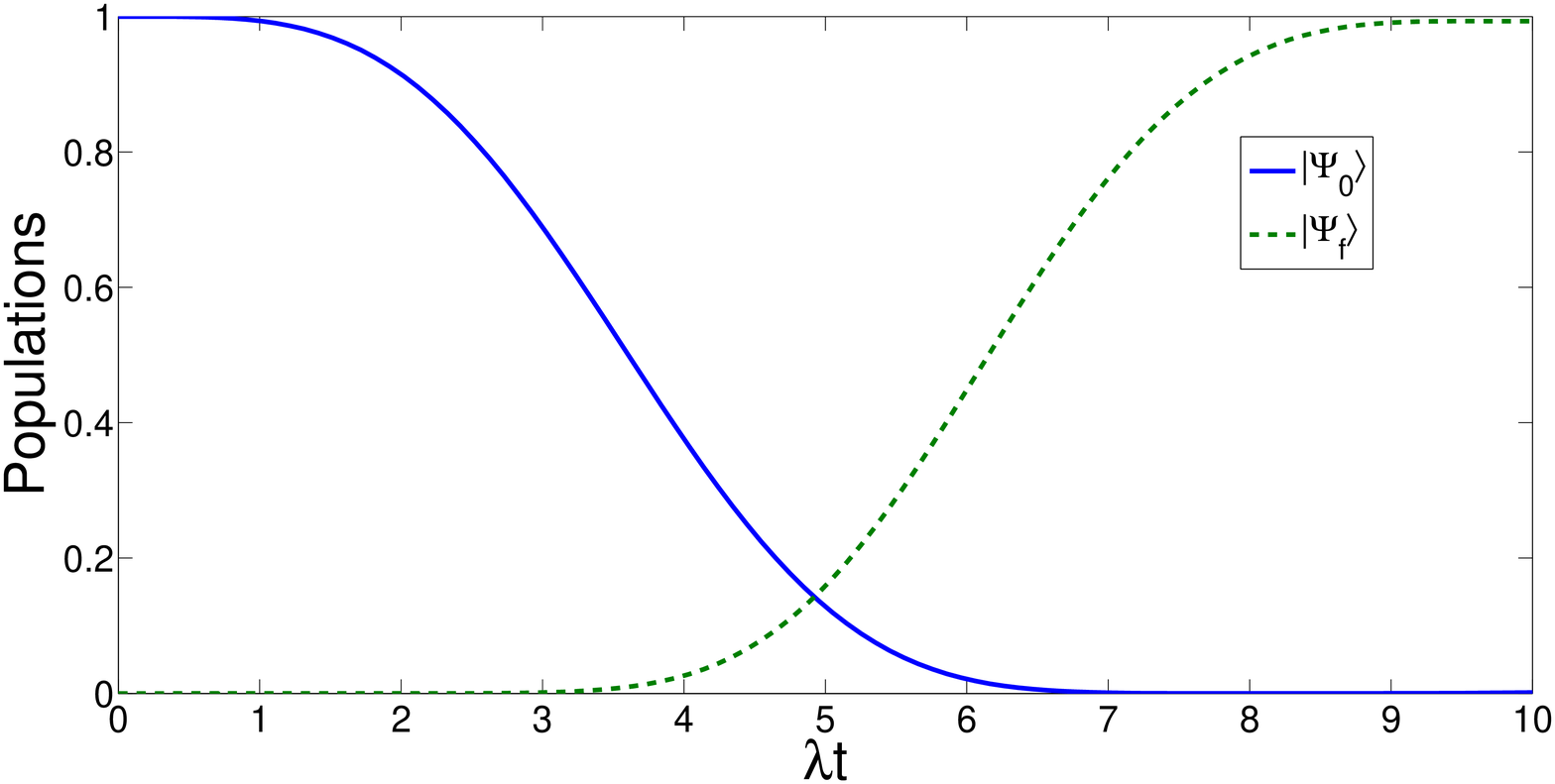}}
 \subfigure[]{
 \includegraphics[scale = 0.19]{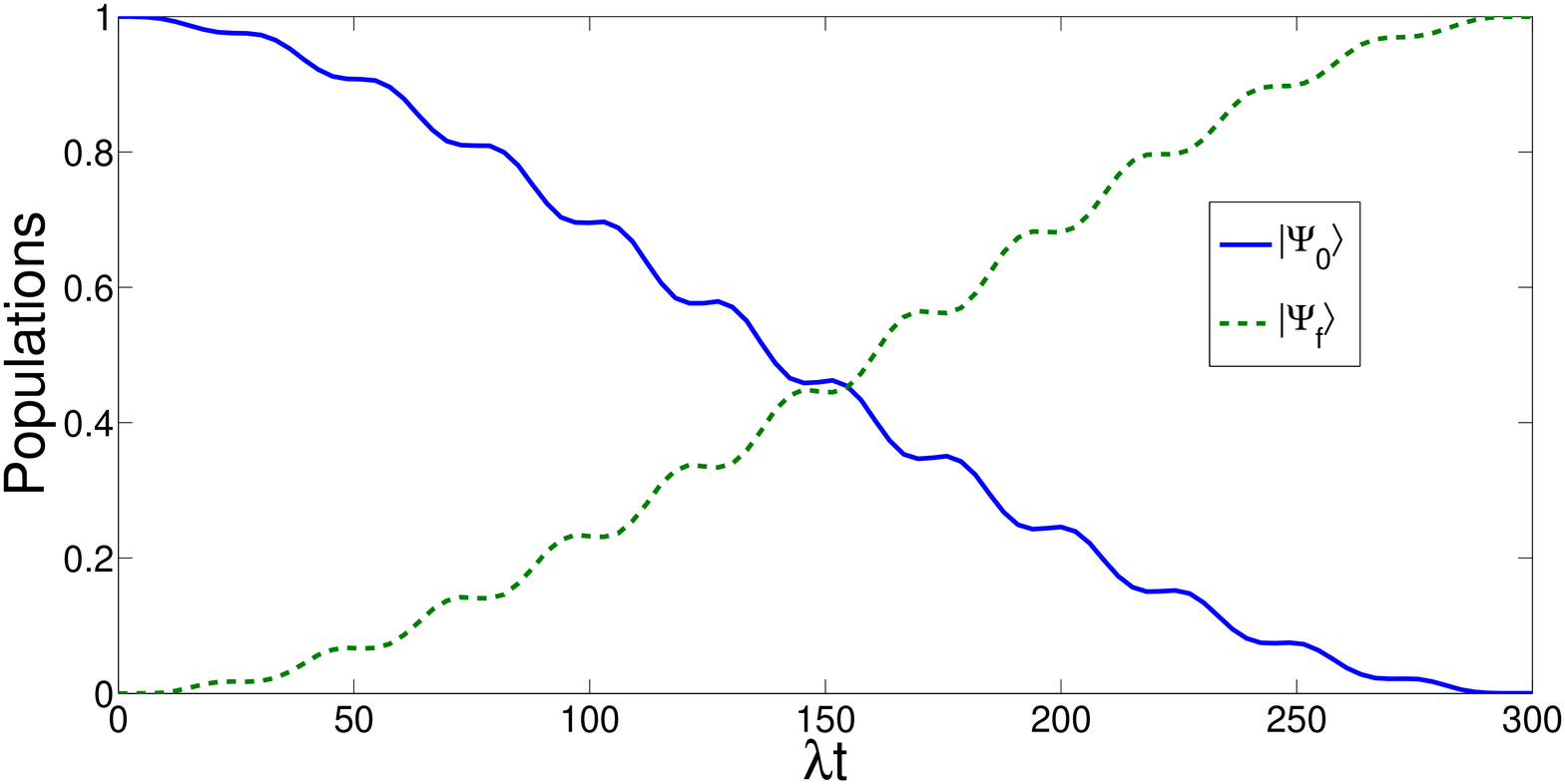}}
 \caption{
    Time-dependent populations in STAP schemes and the comparisons between STAP methods and adiabatic methods:
    (a) Time evolution of the populations for the states $|\Psi_{1}\rangle$, $|\Psi_{f}\rangle$ and $|\Psi_{a}\rangle$ of three-atom GHZ-state via STAP.
    (b) The comparison of total interaction times required between STAP method and adiabatic method for GHZ-state generation.
    (c) Time evolution of the populations for the states $|\Phi_{1}\rangle$, $|w_{1}\rangle$, $|w_{2}\rangle$ and $|w_{3}\rangle$ of three-atom $W$ via STAP.
    (d) The comparison of total interaction times required between STAP method and adiabatic method for $W$-state generation.
    (e) Time evolution of the entanglement transfer form the initial entangled state $|\Psi_{0}\rangle$ to target entangled state $|\Psi_{f}\rangle$ via STAP.
    (f) Time evolution of the entanglement transfer form the initial entangled state $|\Psi_{0}\rangle$ to target entangled state $|\Psi_{f}\rangle$ via adiabatic passage.
          }
\label{GHZW}
\end{figure}

For simplicity, in the following when a comparison between the
adiabatic method and the STAP method should be taken, the form of
the Hamiltonian for an adiabatic method is chosen as the same as
that for a STAP method. Figure. \ref{GHZW} (a) shows the
time-dependent populations for states $|\Psi_{1}\rangle$,
$|\Psi_{f}\rangle$ and $|\Psi_{a}\rangle$ in the scheme of fast GHZ
state generation when $\{\epsilon=\arcsin{0.25},t_{f}=10/\lambda\}$,
and Fig. \ref{GHZW} (b) gives the time comparison between the
adiabatic method and the STAP method. The dashed curve in Fig.
\ref{GHZW} (b) is plotted with parameters
$\{\epsilon=\arcsin{0.02},t_{f}=60/\lambda\}$ satisfying the
adiabatic condition via adiabatic passage. It is obvious that Figs.
\ref{GHZW} (a) and (b) are almost as the same as Figs. \ref{Bell}
(a) and (b), respectively. Similar phenomenon is shown in the
generation of $W$ state when we contrast Figs. \ref{GHZW} (c) and
(d) with Figs. \ref{Bell} (c) and (d). Here, Fig. \ref{GHZW} (c)
displays the populations for states $|\Phi_{1}\rangle$,
$|w_{1}\rangle$, $|w_{2}\rangle$, and $|w_{3}\rangle$ when
$\{\epsilon=\arcsin{0.25},t_{f}=10/\lambda\}$, where
$|w_{1}\rangle=|g\rangle_{p}|f,g,g\rangle_{1,2,3}|0,0,0\rangle_{c_{1},c_{2},f}$,
$|w_{2}\rangle=|g\rangle_{p}|g,f,g\rangle_{1,2,3}|0,0,0\rangle_{c_{1},c_{2},f}$,
and
$|w_{3}\rangle=|g\rangle_{p}|g,g,f\rangle_{1,2,3}|0,0,0\rangle_{c_{1},c_{2},f}$,
and Fig. \ref{GHZW} (d) gives the comparison between STAP method and
adiabatic method for $W$ state generation. We choose parameters
$\{\epsilon=\arcsin{0.02},t_{f}=100/\lambda\}$ to satisfy the
adiabatic condition in the adiabatic method. As a matter of fact,
the STAP methods for GHZ state and $W$ state generations are
equivalent to the first and second STAP methods for Bell-state
generations, respectively. When $M=2$, the STAP methods for
$M$-particle-entangled-state generations are exactly the STAP
methods for the Bell-state generations. From the result of analysis,
a significant inference is obtained: the interaction time required
for each of the generations of GHZ state and $W$ state is irrelevant
to the number of qubits. Whereas, when there are too many atoms
trapped in the cavity $c_{2}$, the system is possibly not safe to
assume linear optic models since the couplings $\lambda_{s}$ is
proportionally reduced while $\Omega_{s}(t)$ remains unchanged in
Sec. \textrm{IV} C if we do not change the parameters. Therefore, in
fact, the shortest interaction time required for $W$ state
generation is limited by the number of qubits. But that only causes
a small effect on the STAP method since even when $M=100$, we can
still choose the interaction time as $t_{f}=50/\lambda$ which is still
much shorter than that in an adiabatic method to make sure
the system is safe to assume linear optic models (in this case,
$\Omega_{0}\approx2\lambda_{s}$). Figure \ref{GHZW} (e) shows the time
evolution of the populations for the initial entangled state
$|\Phi_{0}\rangle$ and target entangled state $|\Psi_{f}\rangle$ in
the scheme of transition of two-atom entangled state when
$\epsilon=\arcsin{0.25}$, $t_{f}=10/\lambda$ and $v=\lambda$. It is
merited that Fig. \ref{GHZW} (e) is as the same as Fig. \ref{p0p7}
(b), the reason for this phenomenon is also that the Hamiltonian in
Eq. (\ref{eq5-4}) is precisely the same to the Hamiltonian in Eq.
(\ref{eq3-1}). And Fig. \ref{GHZW} (f) is the two-atom entanglement
transfer via adiabatic passage in case of
$\{t_{f}=300/\lambda,\epsilon=\arcsin{0.02}\}$.

\begin{figure}
 \renewcommand\figurename{\small FIG.}
 \centering \vspace*{8pt} \setlength{\baselineskip}{10pt}
 \subfigure[]{
 \includegraphics[scale = 0.19]{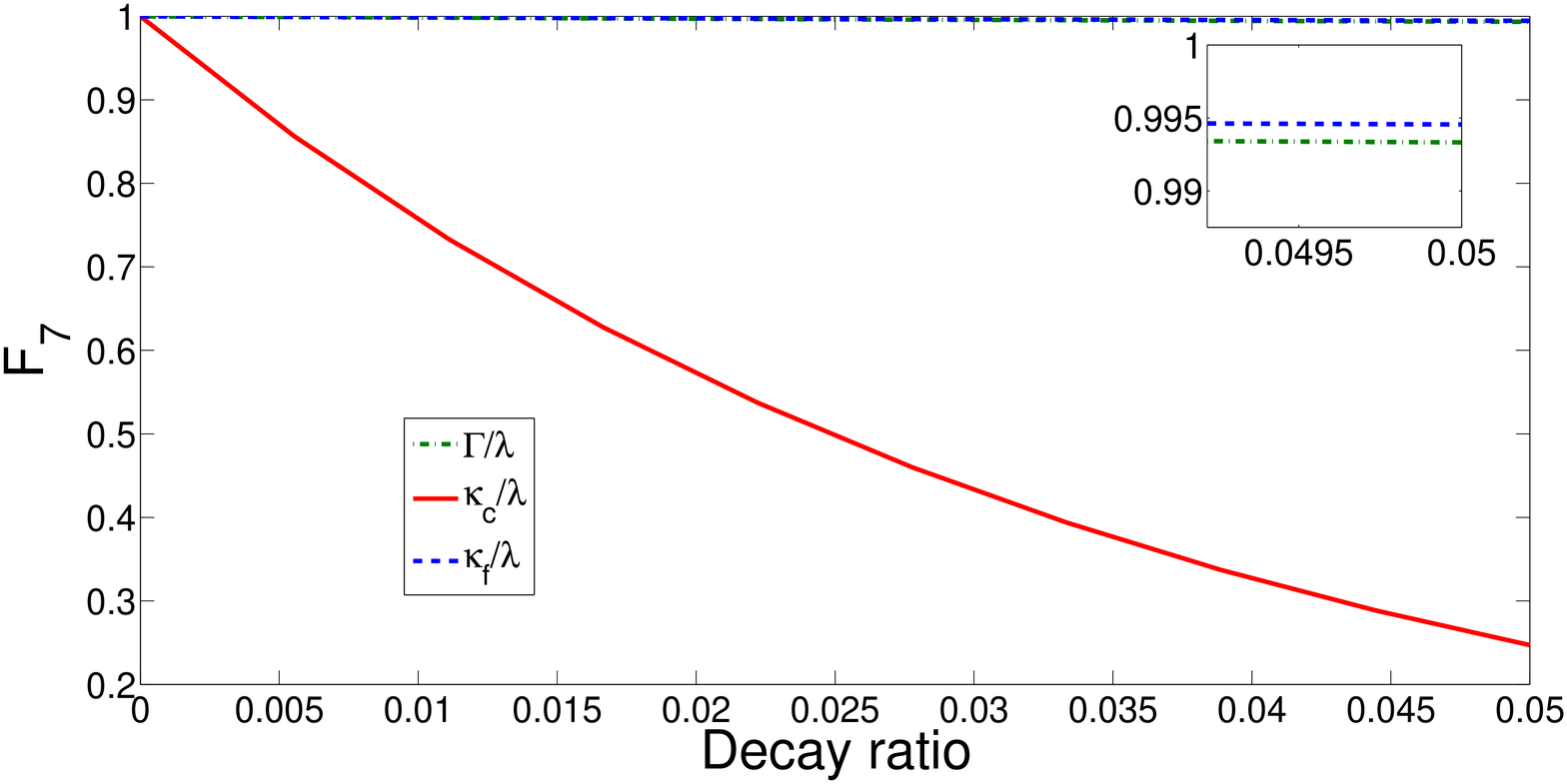}}
 \subfigure[]{
 \includegraphics[scale = 0.19]{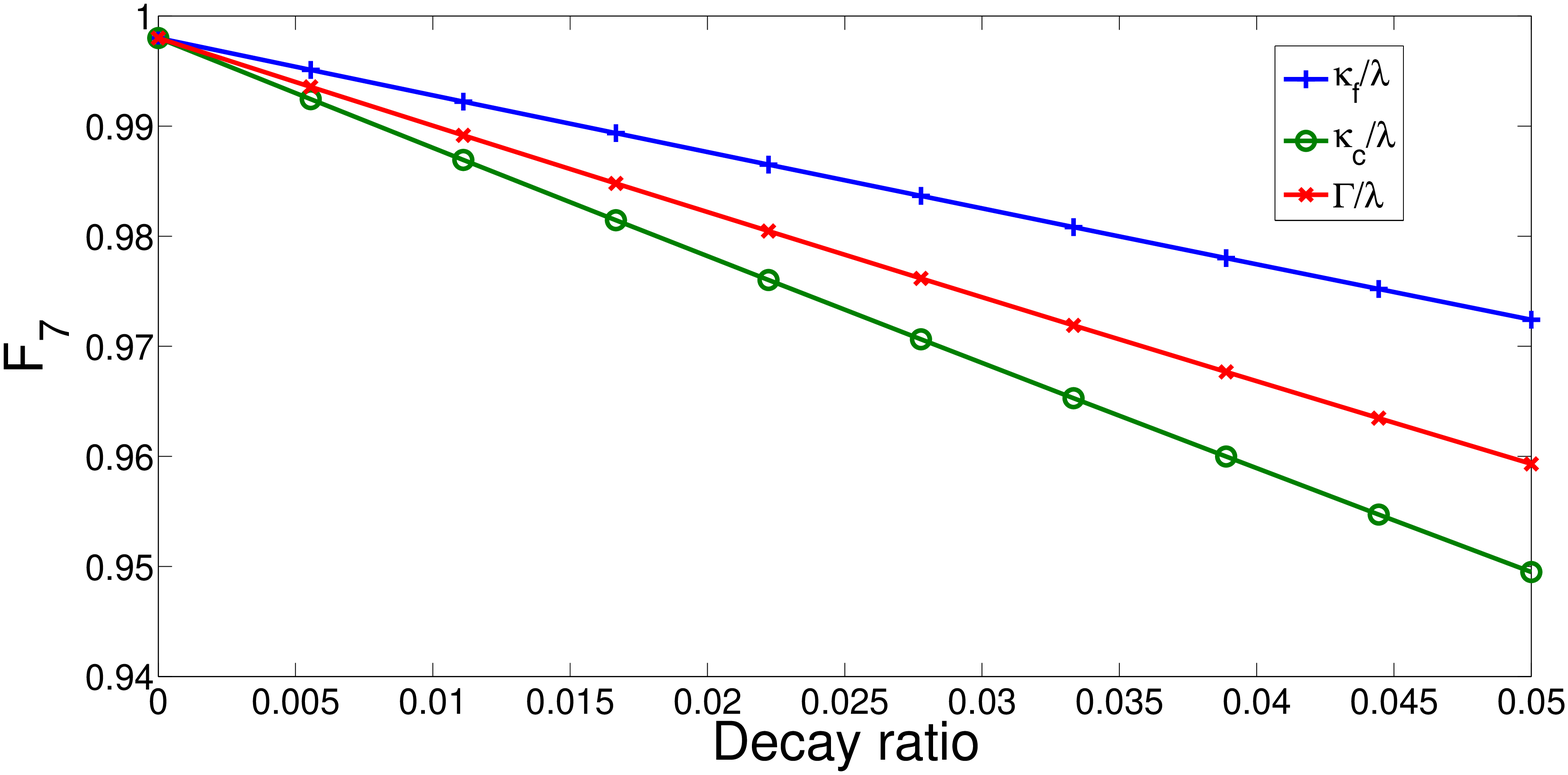}}
 \caption{
    (a) The influence of decays on the fidelity $F_{7}$ of the target state via adiabatic method: each of the three curves denotes $F_{7}$ versus
        each of the three noise resources when the other two are zero,
        for example the blue curve denotes the relationship between the fiber decay $\kappa_{f}/\lambda$ and the $F_{7}$ when $\kappa_{c}=\Gamma=0$.
    (b) The influence of decays on the fidelity $F_{7}$ of the target state via STAP method.
          }
\label{Fkfcr}
\end{figure}

As is known to all, whether a scheme is applicable for quantum
information processing and quantum computing depends on the
robustness against possible mechanisms of decoherence. The
dissipation has not been taken into account in the above discussion.
And once the dissipation is considered, the evolution of the system
can be modeled by a master equation in Lindblad form,
\begin{eqnarray}\label{eq6-2}
  \dot{\rho}=i[\rho,H]+\sum_{k}[L_{k}\rho L_{k}^{\dag}-\frac{1}{2}(L^{\dag}_{k}L_{k}\rho+\rho L_{k}^{\dag}L_{k})],
\end{eqnarray}
where the $L_{k}$'s are the so-called Lindblad operators. For the
first scheme, there are seven Lindblad operators:
\begin{eqnarray}\label{eq6-3}
  L_{1}^{\kappa_{c}}&=&\sqrt{\kappa_{1}}a_{1},
  L_{2}^{\kappa_{c}}=\sqrt{\kappa_{2}}a_{2},
  L_{3}^{\kappa_{f}}=\sqrt{\kappa_{f}}b,   \cr
  L_{4}^{\Gamma}&=&\sqrt{\Gamma_{1}}|f\rangle_{1}\langle e|,
  L_{5}^{\Gamma}=\sqrt{\Gamma_{2}}|g\rangle_{1}\langle e|,
  L_{6}^{\Gamma}=\sqrt{\Gamma_{3}}|f\rangle_{2}\langle e|,
  L_{7}^{\Gamma}=\sqrt{\Gamma_{4}}|g\rangle_{2}\langle e|,
\end{eqnarray}
where $\kappa_{m}$ ($m=1,2$) are the decays of the cavities,
$\kappa_{f}$ is the decay of the fiber, and $\Gamma_{n}$
($n=1,2,3,4$) are the spontaneous emissions of atoms. We assume
$\kappa_{m}=\kappa_{c}$ and $\Gamma_{n}=\Gamma/2$, in the interest
of simplicity. Adiabatic passage technique has an advantage that the
fidelity of the target state is independent of the fiber decay and
atomic spontaneous emission because the intermediate states
including the excited state of the fibers and atoms are always
adiabatically eliminated, and the decoherence is usually caused by
the cavity decay since the dark state always includes the terms
including the cavity-excited states. We prove this by numerical
simulation shown in Fig. \ref{Fkfcr} (a).  Figure \ref{Fkfcr} (a)
displays the fidelity $F_{7}$ versus each of the three noise
resources when the other two are zero via adiabatic process (we
choose parameters $\{\epsilon=0.02,t_{f}=100/\lambda\}$ to satisfy
the adiabatic condition). Figure \ref{Fkfcr} (b) shows the
relationship between $F_{7}$ and the three noise resources via STAP.
Contrast these two figures; the benefits of the STAP method are shown obviously; though the STAP method
is a little more sensitive to the fiber decay and atomic spontaneous
emission, it is far more robust again the cavity decay than the
adiabatic method. In a word, the STAP method is not only fast but
also robust.

\begin{figure}
 \renewcommand\figurename{\small FIG.}
 \centering \vspace*{8pt} \setlength{\baselineskip}{10pt}
 \subfigure[]{
 \includegraphics[scale = 0.19]{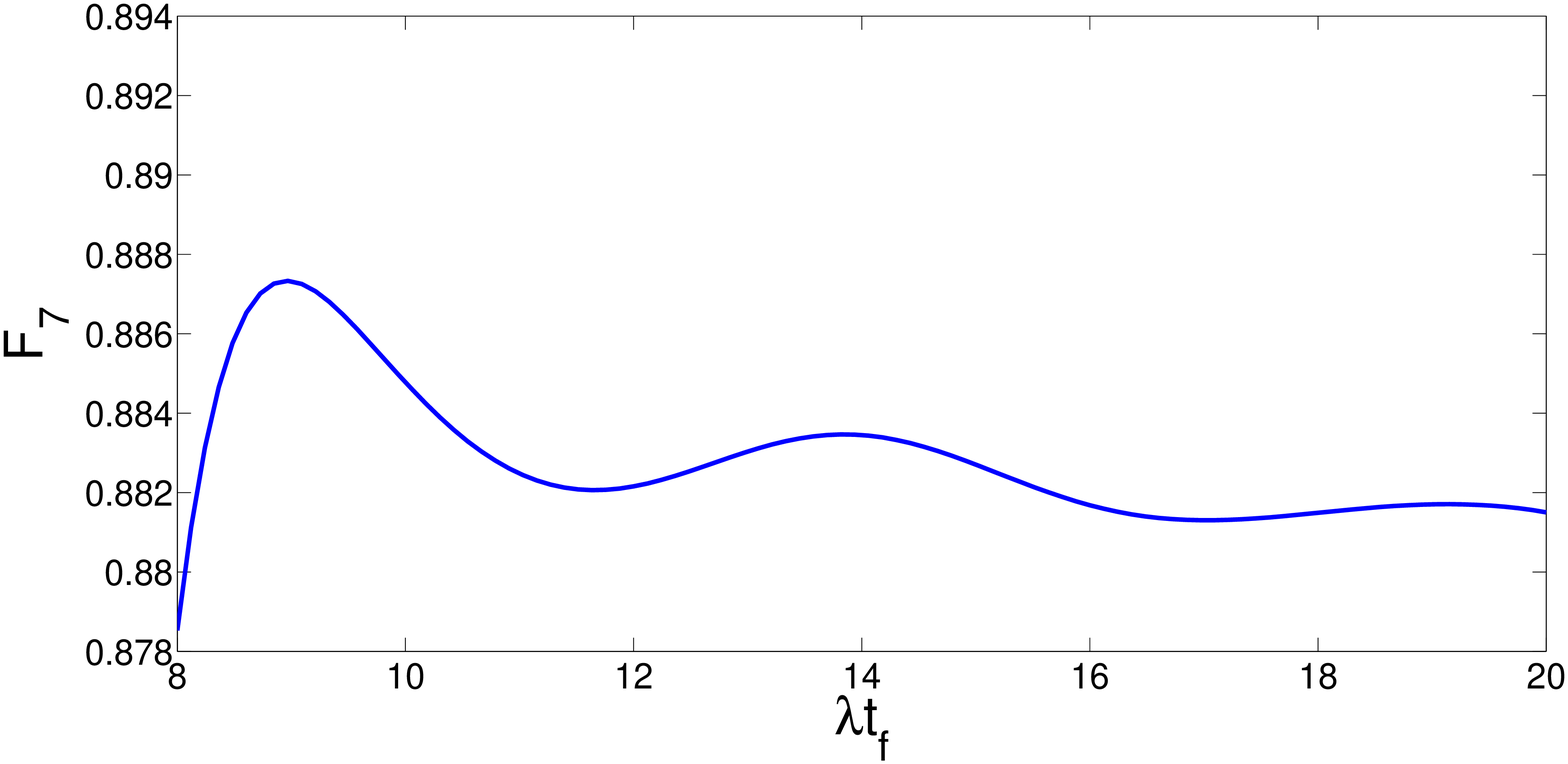}}
 \subfigure[]{
 \includegraphics[scale = 0.19]{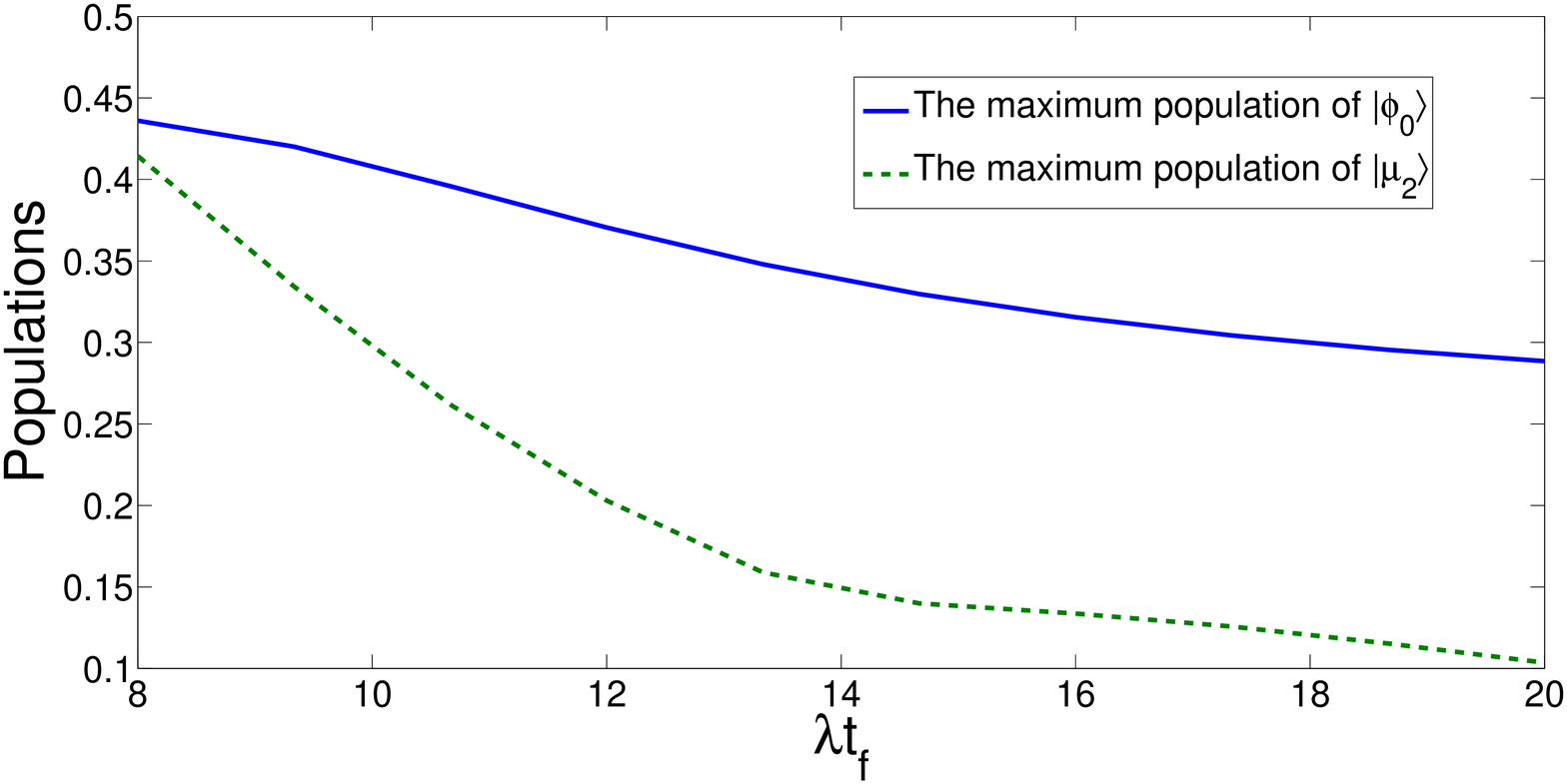}}
 \caption{
    (a) The relationship between the fidelity of the target state $|\psi_{7}\rangle$ and total interaction time $\lambda t_{f}$ in the presence of decoherence.
    (b) The maximum populations of states $|\phi_{0}\rangle$ and $|\mu_{2}\rangle$ versus the total interaction time $\lambda t_{f}$.
          }
\label{Ftf}
\end{figure}

\begin{figure}
 \renewcommand\figurename{\small FIG.}
 \centering \vspace*{8pt} \setlength{\baselineskip}{10pt}
 \subfigure[]{
 \includegraphics[scale = 0.19]{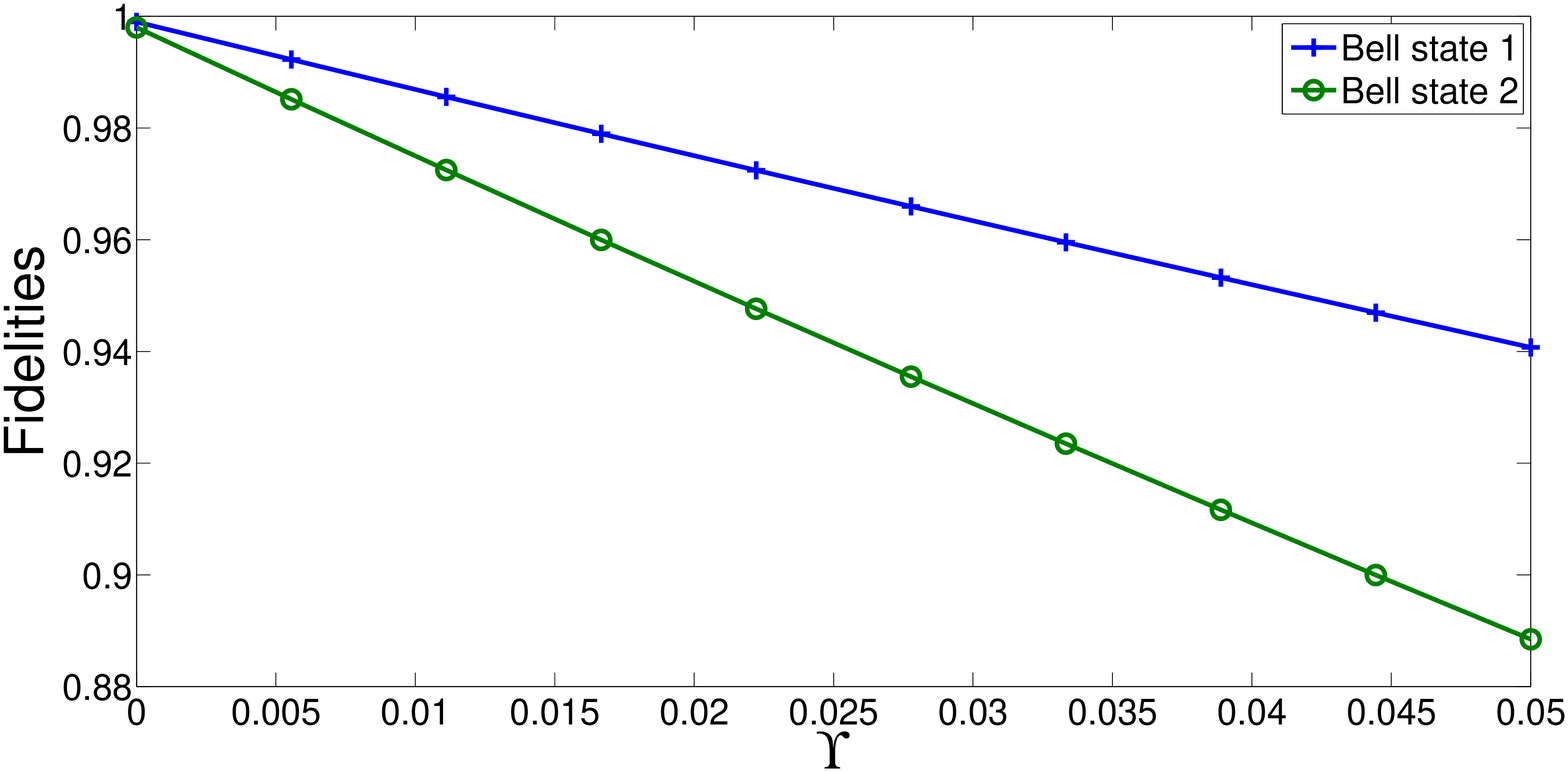}}
 \subfigure[]{
 \includegraphics[scale = 0.19]{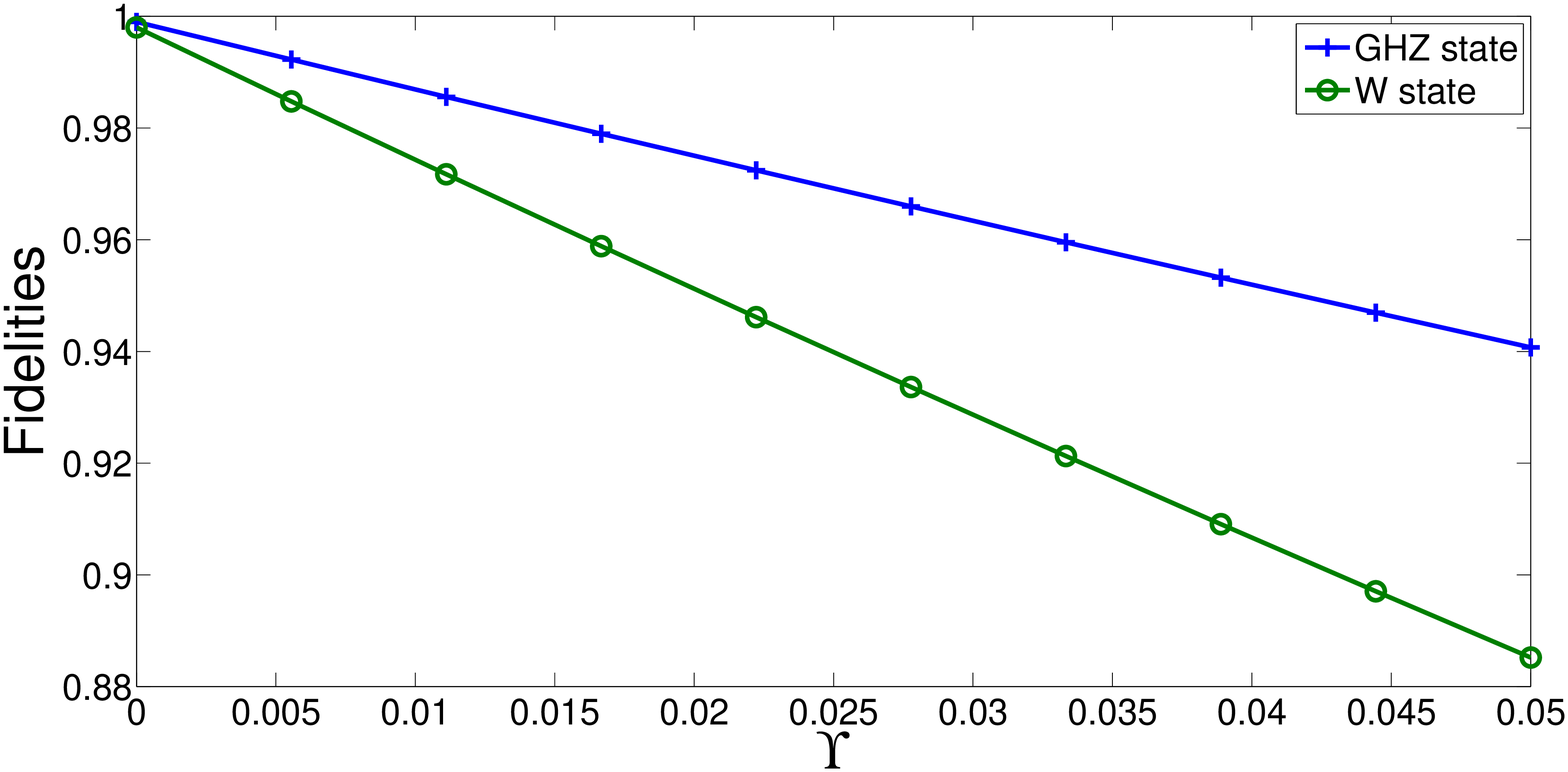}}
 \caption{
    (a) The fidelities of the Bell states versus the the ratios of decays by choosing $\Gamma/\lambda=\kappa_{f}/\lambda=\kappa_{c}/\lambda=\Upsilon$.
    (b) The fidelity of the three-atom GHZ state and the fidelity of the three-atom $W$ state versus the the ratios of decays by choosing
    $\Gamma/\lambda=\kappa_{f}/\lambda=\kappa_{c}/\lambda=\Upsilon$.
          }
\label{entangledkfcr}
\end{figure}

The decoherence, as a general rule, is dominated by the populations
of the excited states and the total operation time. In the previous
articles, the decoherence is always refrained by decreasing the
populations of the excited sates. However, the excited states are
required to speed up the transfer in the present STAP method.
Therefore, we plot Fig. \ref{Ftf} (a) to demonstrate that in the
present case, the decoherence is mainly dominated by the total
operation time rather than the populations of the excited states.
The result obviously shows that the fidelity is the highest when the
operation time for achieving an ideal target state is the shortest
i.e., $t_{f}=8.9/\lambda$. And the fidelity decreases gradually as
$\lambda t_{f}$ increases while oscillating. When
$t_{f}<8.9/\lambda$, as the conditions for constructing STAP are no
longer faultlessly satisfied, the fidelity is lower than that when
$t_{f}=8.9/\lambda$. The maximum populations of the intermediate
states $|\phi_{0}\rangle$ and $|\mu_{2}\rangle$ which practically
cause the decoherence versus $\lambda t_{f}$ are shown in Fig.
\ref{Ftf} (b). The maximum populations clearly decrease with the
increasing of $\lambda t_{f}$. Join up these two results, an
inference is drawn, that is, at least when $9/\lambda\leq t_{f}\leq
20/\lambda$, the decoherence is mainly dominated by the total
operation time since decreasing the populations of the excited sates
does not obviously refrain the decoherence.
\ Figure \ref{Ftf} (a) is plotted with parameters
$\epsilon=\arcsin{0.25}$, $v=g$, and
$\kappa_{c}=\kappa_{f}=\Gamma=0.05\lambda$, and Fig. \ref{Ftf} (b)
is plotted with parameters $\epsilon=\arcsin{0.25}$, $v=g$, and
$\kappa_{c}=\kappa_{f}=\Gamma=0$,

The robustness of the STAP for the Bell-state generations is shown
in Fig. \ref{entangledkfcr} (a) with
$\Gamma/\lambda=\kappa_{f}/\lambda=\kappa_{c}/\lambda=\Upsilon$. And
when we trap one more atom in the cavities, two more Lindblad
operators $L_{n}^{\Gamma}=\sqrt{\Gamma_{n}}|S\rangle_{n}\langle e|$
($S=f,g$) governing spontaneous emissions should be considered in
the master equation. The result proves that the first STAP method
for Bell-state generation is much more robust against decoherence
than the second one. That is, because only the term
$\frac{1}{\sqrt{2}}|\psi_{1}\rangle$ in Eq. (\ref{eq4a-1})
participates in the evolution. From such an initial state to the
target state $\frac{1}{\sqrt{2}}|\psi_{7}\rangle$ along the STAP,
the evolution is very closely analogous to the evolution of the FPT
scheme in Sec. \textrm{III} except all of the coefficients of the
states should be multiplied by $\frac{1}{\sqrt{2}}$. However, the
second scheme of Bell-state generation is actually equivalent to the
scheme for FPT. That means the population for each of the effective
excited states in the second scheme is twice as large as that in the
first scheme at any time during the evolution. Figure
\ref{entangledkfcr} (b) displays the fidelities of the three-atom
GHZ and $W$ states versus the decays when
$\Gamma/\lambda=\kappa_{f}/\lambda=\kappa_{c}/\lambda=\Upsilon$.
Apparently Fig. \ref{entangledkfcr} (b) is almost identical with
Fig. \ref{entangledkfcr} (a). Further research shows that, when it
comes to the generations of more qubits GHZ and $W$ states, the
robustness of the schemes against decoherence still remains the same
as that of three atoms. It is implied that not only the interaction
time required for the generation of entangled states but also the
robustness of the scheme against decoherence is irrelevant to the
number of qubits. The reason for this phenomenon is not hard to
understand: however much atoms are trapped in the cavities, the
photon emitted from the atom $1$ in the cavity $c_{1}$ can only
excite one of atoms in cavity $c_{2}$. In other words, the holistic
spontaneous emissions of $M$ atoms is equivalent to that of one atom
in the cavity $c_{2}$. The result applies to the scheme of the
transition of entangled states as well.

Finally, we present a brief discussion about the basic elements in
the real experiment. The cesium atoms can be used to implement the
schemes, the state $|g\rangle$ corresponds to $F=3$, $m=2$ hyperfine
state of $6^{2}S_{1/2}$ electronic ground state, $|a\rangle$
corresponds to $F=3$, $m=4$ hyperfine state of $6^{2}S_{1/2}$
electronic ground state, $|f\rangle$ corresponds to $F=4$, $m=3$
hyperfine state of $6^{2}S_{1/2}$ electronic ground state, and
$|e\rangle$ corresponds to $F=4$, $m=3$ hyperfine state of
$6^{2}P_{1/2}$ electronic state. An almost perfect fiber-cavity
coupling with an efficiency larger than $99.9\%$ can be realized
using fiber-taper coupling to high-Q silica microspheres
\cite{SMSTJKOJPKJVPrl03}. The fiber loss at $852$-nm wavelength is
only about $2.2$ dB/Km \cite{KJGVFPDTGSBIeee04}, in this case, the
fiber decay rate is only $0.152$ MHz. While in recent experimental
conditions \cite{JRBHJKPra03,SMSTJKKJVKWGEWHJKPra05}, it is
predicted to achieve a strong atom-cavity coupling
$\lambda=2\pi\times750$ MHz. That means the fiber decay can actually
be neglected in a real experiment with these parameters. And by
choosing another set of parameters
$(\lambda,\Gamma,\kappa_{c})=(2500,10,10)$ MHz in the microsphere
cavity QED experiment reported in
\cite{SMSTJKKJVKWGEWHJKPra05,MJHFGSLBMBPNat06}, the fidelities of
all of the STAP schemes in this paper are higher than $99\%$.

Till now, all the above description and discussion are based on
atoms are trapped in two distant cavities which are connected by a
fiber, in fact, the present STAP can also be generalized to the
system with $N$ ($N\geq2$) atoms respective trapped in $N$ cavities
which are connected by $N-1$ fibers \cite{SYHYXJSNBAJosab13}. In
such a system, there exist only one eigenvector whose eigenvalue is
zero for the intermediate Hamiltonian and one dark state. Similar to
 Eqs. (\ref{eq3-3} - \ref{eq3-7}) in Sec. III, in
light of quantum Zeno subspace division, we can rewrite the
interaction Hamiltonian with some special vectors and partition the
intermediate states into parts which are independent of each other.
After discarding most of the parts except $S_{0}$ and $S_{1}$ whose
eigenvalues are Zero and closest to zero, respectively, the
effective Hamiltonian which has the same form as that in Eq.
(\ref{eq3-7}) is easily obtained, and then the STAP can also be
constructed to realize the fast and noise-resistant population
  transfer, quantum entangled states preparation, and quantum entangled
  states transition in multi-cavity-fiber-atom combined system.

\section{Conclusion}
``Shortcuts to adiabatic process'' provides alternative fast and
robust processes which reproduce the same final populations, or even
the same final state, as the adiabatic process in a finite, shorter
time. In general, the operation time for a method is the shorter the
better, otherwise, the method may be useless. Many experiments of
quantum information science also desire fast and robust theoretical
methods since high repetition rates contribute to the achievement of
better signal-to-noise ratios and better accuracy. Therefore, in
this paper, through dividing the whole system into parts, we
construct shortcuts in different parts in the same time, and
implement the fast quantum information processing via the STAP.
Different schemes are proposed to perform fast population transfer,
entanglement states generations, and entangled states transition.
Through using the STAP, the problem of proposing effective schemes
which are not only fast, but also robust to perform population
transfer, prepare entanglements and implement entangled states
transition in multi-cavity-fiber-atom combined system have been
successfully overcome. Numerical simulation demonstrates the all of
schemes are fast and robust versus variations in the experimental
parameters and decoherence. Moreover, similar operations can also be
applied to other cavities QED systems, for example,
multi-cavity-cavity coupled system, to realize fast and
noise-resistant quantum information processing. We believe that the
present work is promising and will make great contributions to the
experimental research in  quantum information science. ($\sim93\%$)

\section{ACKNOWLEDGEMENT}

 This work was supported by the National Natural Science Foundation of China under
Grants No. 11105030 and No. 11374054, the Major State Basic Research
Development Program of China under Grant No. 2012CB921601, and the
Foundation of Ministry of Education of China under Grant No. 212085.

\newpage

\end{document}